\documentclass[11pt]{article}

\usepackage[margin=1in]{geometry}
\usepackage{microtype}
\usepackage{amsmath}
\usepackage{booktabs}
\usepackage{graphicx}
\usepackage{xcolor}
\usepackage{float}
\usepackage{hyperref}
\usepackage{tikz}
\usetikzlibrary{arrows.meta,positioning,shapes.geometric,backgrounds,calc,shapes.symbols,patterns}
\usepackage{pgfplots}
\pgfplotsset{compat=1.18}
\usepgfplotslibrary{fillbetween}
\usepackage{listings}
\usepackage{cite}
\usepackage{tabularx}
\usepackage{array}
\usepackage{multirow}
\usepackage{makecell}
\newcolumntype{Y}[1]{>{\raggedright\arraybackslash}p{#1}}
\newcolumntype{L}{>{\raggedright\arraybackslash}X}
\newcolumntype{Z}{>{\raggedright\arraybackslash}X}
\newcommand{\nimmodeid}{\texttt{meta/\allowbreak{}llama-4-\allowbreak{}maverick-\allowbreak{}17b-\allowbreak{}128e-\allowbreak{}instruct}}

\lstdefinelanguage{json}{
  basicstyle=\ttfamily\footnotesize,
  numbers=left,
  numberstyle=\tiny\color{gray},
  stepnumber=1,
  numbersep=6pt,
  xleftmargin=15pt,
  linewidth=\dimexpr\linewidth-15pt\relax,
  showstringspaces=false,
  breaklines=true,
  frame=lines,
  literate=
    *{0}{{{\color{blue!60!black}0}}}{1}
     {1}{{{\color{blue!60!black}1}}}{1}
     {2}{{{\color{blue!60!black}2}}}{1}
     {3}{{{\color{blue!60!black}3}}}{1}
     {4}{{{\color{blue!60!black}4}}}{1}
     {5}{{{\color{blue!60!black}5}}}{1}
     {6}{{{\color{blue!60!black}6}}}{1}
     {7}{{{\color{blue!60!black}7}}}{1}
     {8}{{{\color{blue!60!black}8}}}{1}
     {9}{{{\color{blue!60!black}9}}}{1}
     {:}{{{\color{red!60!black}{:}}}}{1}
     {,}{{{\color{red!60!black}{,}}}}{1}
     {\{}{{{\color{purple}{\{}}}}{1}
     {\}}{{{\color{purple}{\}}}}}{1}
     {[}{{{\color{purple}{[}}}}{1}
     {]}{{{\color{purple}{]}}}}{1}
}

\hypersetup{
  colorlinks=true,
  linkcolor=black,
  urlcolor=blue,
  citecolor=black,
  pdftitle={FuzzPilot: Plateau-Triggered Recipe Validation for Structured Text Fuzzing},
  pdfauthor={Zhiyi Yao},
  pdfsubject={Greybox fuzzing, AFL++, plateau-triggered control, recipe validation, cJSON},
  pdfkeywords={fuzzing, AFL++, LLM, mutation recipe, plateau detection, micro-campaign, Ghidra, cJSON},
}

\newcommand{\circled}[1]{\tikz[baseline=(C.base)]{%
  \node[draw,circle,inner sep=0.5pt,minimum size=2.6mm,
        fill=red!75!black,text=white,font=\tiny\bfseries,
        line width=0.3pt](C){#1};}}

\title{FuzzPilot: Plateau-Triggered Recipe Validation\\
  for Structured Text Fuzzing}

\author{Zhiyi Yao\\
  Qingdao University of Technology\\
  \texttt{202202030221@stu.qut.edu.cn}}

\date{May 24, 2026}

\begin{document}
\maketitle

\begin{abstract}
Greybox fuzzers leave little room for expensive decision making in
the mutation loop. FuzzPilot is a small controller around AFL++ that
keeps those decisions off the hot path. At a coverage plateau, it
takes a corpus snapshot, prepares candidate mutation ``recipes'',
tests them in short isolated AFL++ runs, and promotes a recipe only
when the validation run produces a positive reward. Recipes are data,
not generated code: a native custom mutator consumes a compact JSON
strategy containing operator weights, byte ranges, corpus-selection
rules, and dictionary tokens. Candidate recipes can come from local
rules or from a language-model agent, while Ghidra-derived constants
and decompiled context provide target-specific hints.

This first public report is deliberately narrow. We evaluate the
implementation on cJSON with $N{=}5$ repetitions for the two main
arms, \texttt{baseline-afl} and \texttt{full-agent}, each with a
$14{,}400$\,s budget. cJSON turns out to be a saturated target:
vanilla AFL++ reaches the exposed 269-edge ceiling at a median of
$2{,}524$\,s (time to first reach 269 edges; the distinct
plateau-duration metric used in the next paragraph is $2{,}532$\,s).
As a result, these experiments do not establish
that language-model proposals improve coverage, nor do they support a
generalization claim beyond cJSON.

Within that scope, the system has three useful properties. First,
moving the controller off the hot path preserves throughput:
\texttt{full-agent} has a median \texttt{execs\_per\_sec} about
$1.06\times$ the baseline. Second, \texttt{full-agent} has a
descriptively shorter median plateau, $1{,}384$\,s versus
$2{,}532$\,s for baseline, but the difference is not statistically
significant at $N{=}5$ (Mann--Whitney $p{=}0.42$,
$A_{12}{=}0.68$). Third, the validation gate behaved conservatively:
it evaluated 20 model-proposed recipes and promoted none, because all
candidate rewards were zero on the saturated corpus. Preliminary
ablations suggest that the observed plateau reduction is more likely
due to the controller's snapshot and restart machinery than to the
recipe mutator or the model layer; a dedicated \texttt{controller-only}
arm is left for the larger evaluation. This preprint is therefore best
read as an auditable implementation report and a baseline for the
ongoing non-saturated-target study, not as a claim that every component
already improves fuzzing performance.
\end{abstract}

\section{Introduction}
\label{sec:intro}

Greybox fuzzers such as AFL++ are built around a simple constraint:
the mutation loop must stay fast. On small parsers, a useful campaign
can execute hundreds of thousands of test cases per second, so even a
modest amount of per-input reasoning can dominate the run time. This
constraint is awkward for systems that want to use richer target
signals, such as decompiled constants, format-specific tokens, or
language-model suggestions. Such signals are often useful, but they do
not belong in the inner loop.

Recent LLM-assisted fuzzers explore several ways to manage that cost.
Some invoke the model as an input generator or mutator during the
campaign~\cite{wang2024llamafuzz,liu2024fuzzcoder,xia2024fuzz4all};
others move model work into a helper process~\cite{lin2025hybrid}.
G\textsuperscript{2}FUZZ~\cite{liu2025g2fuzz} goes further by asking
the model for a reusable Python generator only when the fuzzer stops
making progress. That line of work suggests a practical rule for
fuzzer design: expensive reasoning should happen at coarse intervals,
and the product of that reasoning should be something the fuzzer can
reuse cheaply.

FuzzPilot follows that rule for structured-text parsers. Instead of
asking for executable generators, it works with mutation recipes:
small data objects that describe how a native AFL++ custom mutator
should bias its operators, byte ranges, token choices, and seed
selection. Recipes can be written by local rules or proposed by a
language-model agent. Either way, FuzzPilot does not trust a candidate
immediately. At a plateau, it snapshots the current queue, tests
candidate recipes in short AFL++ micro-campaigns, and promotes only a
candidate that earns a positive reward. The design is intentionally
conservative: an unhelpful candidate should cost a bounded validation
run, not poison the main campaign.

\paragraph{Contributions.} This report contributes the implemented
FuzzPilot design and an initial audit of its behavior. The claims are
intentionally limited: cJSON exercises the machinery end to end, but
because the target saturates early, it does not show that every
component is beneficial.

\paragraph{Scope and limitations (stated up front).} Before
enumerating contributions, we make the bounds of this preprint
explicit. The limits are part of the result, not footnotes:

\begin{itemize}
\item \textbf{Single saturated target}: We evaluate on \emph{only}
cJSON, at $N{=}5$ repetitions for main arms and $N{=}3$ for
ablations. cJSON is a \emph{saturated target} where baseline AFL++
alone exhausts the reachable 269-edge ceiling at a median of
$2{,}524$\,s (time to first reach 269 edges; the distinct
plateau-duration metric used elsewhere is $2{,}532$\,s), leaving no
headroom for the LLM proposal layer or micro-campaign gate to
demonstrate coverage gains. We make \emph{no} empirical claim about
libpng, libxml2, sqlite3, openssl, or any other target.

\item \textbf{LLM layer untested}: The micro-campaign gate evaluated
20 LLM-proposed recipes on cJSON and promoted zero, because target
saturation leaves reward $R{=}0$ on every candidate
(\S\ref{sec:eval:rq3}). The gate's intended property---distinguishing
good recipes from bad on non-saturated targets---remains
\emph{empirically undemonstrated}. A non-saturated-target evaluation
is the most important single piece of future work.

\item \textbf{Statistical power insufficient}: At $N{=}5$ (main) and
$N{=}3$ (ablations), no pairwise plateau or coverage difference
reaches conventional significance thresholds ($p{<}0.05$). We report
descriptive point estimates and effect sizes ($A_{12}$), not
significance claims. Formal rank-sum tests require $N{\geq}20$ for
main arms and $N{\geq}10$ for ablations~\cite{klees2018evaluating}.

\item \textbf{Attribution incomplete}: Ablation experiments
(\S\ref{sec:eval:rq4-ablation}) suggest the plateau reduction is
driven primarily by the controller's plateau-detection and
corpus-reorganization machinery, not by the recipe-guided mutator or
LLM layer. However, a \texttt{controller-only} ablation (plateau
detector + corpus snapshot only, no mutator, no LLM, no Ghidra) is
\emph{not} in the preprint matrix; without it we cannot fully
separate ``controller machinery'' from ``default rule recipe''. This
ablation is deferred to the venue version.

\item \textbf{Fairness baseline caveat}: An E4 fairness arm shows
that AFL++ with \texttt{cmplog} (AFL++'s comparison-operand logging
pass) instrumentation reaches 270 edges (vs FuzzPilot's 269) without
shortening the plateau (\S\ref{sec:eval:rq4-ablation}), indicating
the two techniques operate on complementary axes (wall-time vs
absolute coverage).
\end{itemize}

Taken together, these constraints mean that this version should be
read as an implementation report and a cJSON proof-of-concept. The
larger non-saturated-target matrix, higher repetition counts, and the
missing \texttt{controller-only} ablation are the next steps.
\S\ref{sec:limits} gives the full list of threats to validity. With
these bounds explicit, the contributions are:

\begin{enumerate}
  \item[\textbf{C1}] A \textbf{data-as-recipe} intermediate
  representation that replaces generated helper code with a
  schema-validated JSON description of mutation strategy ---
  a weighted distribution over seven mutation operators, plus
  focus/protect byte ranges, dictionary tokens, and a corpus
  selector --- dispatched by a native AFL++ custom mutator. This
  side-steps the runtime-safety concern of executing generated
  code, makes proposals cacheable and version-able, and lets the
  mutator hot path remain pure native code at AFL++ speed
  (\S\ref{sec:recipe}).
  \item[\textbf{C2}] A \textbf{plateau-triggered controller} that
  detects coverage stalls via a moving-window edge-discovery rate,
  takes a corpus snapshot at plateau time, and can trigger
  intervention strategies (model proposals, corpus reorganization, or
  rule-based recipes). The controller operates off the hot path and
  does not block AFL++ workers. Ablation experiments
  (\S\ref{sec:eval:rq4-ablation}) suggest this machinery is the
  primary driver of the observed plateau reduction on cJSON, though
  a \texttt{controller-only} ablation to isolate its contribution is
  deferred to the venue version (\S\ref{sec:limits}).
  \item[\textbf{C3}] A \textbf{micro-campaign validation gate} that
  scores each candidate recipe in an $N$-second isolated AFL++ run
  over a corpus snapshot and promotes only those with positive
  reward (new edges, paths, or crashes). The gate gives the
  controller empirical evidence about each proposal before it enters
  the main loop. On cJSON, the gate evaluated 20 LLM-proposed
  recipes and promoted zero due to target saturation
  (\S\ref{sec:eval:rq3}); its intended property---distinguishing
  good recipes from bad---remains untested and requires a
  non-saturated target.
  \item[\textbf{C4}] A \textbf{static-analysis context channel}
  built on Ghidra headless that extracts magic tokens, comparison
  constants, branch constraints, and decompiled C from stripped
  binaries. The signal feeds both the agent blackboard (for LLM
  prompts, when enabled) and the AFL dictionary (for direct mutation guidance).
  The ablation E2c (\S\ref{sec:eval:rq4-ablation}) jointly
  quantifies the loss of both consumers; isolating Ghidra's
  advantage over source-level alternatives is deferred to the venue
  paper (\S\ref{sec:limits}).
\end{enumerate}

The artifact also includes an auditable decision trail covering
blackboard state, proposal generation, recipe emission,
micro-campaign reward, and promotion outcome, plus native
Linux/x86\_64 run metadata from the experiment host.

\paragraph{Roadmap.} \S\ref{sec:bg} surveys the design space and
positions FuzzPilot relative to G\textsuperscript{2}FUZZ and the
in-loop line of work. \S\ref{sec:design} presents the architecture.
\S\ref{sec:recipe} details the recipe representation and contrasts
data-as-recipe with code-as-generator. \S\ref{sec:plateau} specifies
the plateau detector, the Ghidra context channel, and the
micro-campaign protocol. \S\ref{sec:eval} reports preliminary
results on cJSON. The per-run decision trail (events log, recipe
proposals, micro-campaign reward, promotion outcome) is recoverable
from \texttt{results/\ldots/runs/p1\_e1\_cjson\_full-agent\_r*/events.jsonl}.
\S\ref{sec:related} relates
FuzzPilot to the broader LLM-fuzzing literature.
\S\ref{sec:limits} states limitations and threats to validity.

\section{Background and the Off-Hot-Path Paradigm}
\label{sec:bg}

\subsection{The AFL++ mutation loop and its throughput budget}
\label{sec:bg:afl}

AFL++ implements a deterministic-then-havoc mutation
loop~\cite{aflplusplus}: each cycle takes an input from the queue,
applies a sequence of bit/byte mutations, executes the target via
\texttt{fork+execve} or a persistent harness, and updates the
coverage bitmap. On small parsers such as cJSON our 4-core x86\_64
host sustains roughly $13{,}000$ executions per second per worker.
Any redesign that interposes a non-trivial computation on this loop
caps end-to-end throughput at the rate of that computation:
inserting a single $100\,\mathrm{ms}$ LLM call per mutation drops
throughput by three orders of magnitude. Existing
LLM-in-the-loop designs accept this cost in exchange for semantic
input quality.

\subsection{In-loop LLM fuzzers}
\label{sec:bg:inloop}

LLAMAFUZZ~\cite{wang2024llamafuzz} fine-tunes an LLM on seed-mutation
pairs and queries it as part of the mutation step; the model output
is then handed to AFL++. FuzzCoder~\cite{liu2024fuzzcoder} casts
mutation as a sequence-to-sequence task and predicts mutation
positions and strategies per input.
Fuzz4All~\cite{xia2024fuzz4all} drives an LLM as the primary input
generator and mutation engine across multiple input languages.
The hybrid design of Lin et al.~\cite{lin2025hybrid} reduces
the per-mutation cost by isolating LLM work in a helper fuzzer at
roughly $250$ queries per hour, with a syntax-validator-and-repair
filter, while a master fuzzer runs AFL++ at full speed. None of
these designs remove the LLM from steady-state fuzzing.

\subsection{The off-hot-path paradigm}
\label{sec:bg:offhot}

G\textsuperscript{2}FUZZ~\cite{liu2025g2fuzz} invokes the model
\emph{only} when the underlying fuzzer fails to find new coverage,
and it does so by asking the model to synthesize a Python input
generator. The generator is executed to produce a batch of candidate
inputs, which AFL++ then mutates as usual. The model contributes
strategy at coarse temporal granularity; mutation throughput remains
native. G\textsuperscript{2}FUZZ shows this paradigm is practical
on $34$ non-textual input formats and outperforms AFL++,
Fuzztruction, and FormatFuzzer in coverage and bug-finding.
We treat the off-hot-path placement as a starting \emph{contract}
that FuzzPilot inherits rather than a contribution we claim.

\subsection{Why structured text needs a different intermediate}
\label{sec:bg:text}

G\textsuperscript{2}FUZZ's intermediate representation is
\emph{Python code}: the model emits a generator script and the
fuzzer executes it. Two properties of the non-textual binary
setting make this choice attractive: (i) binary formats have rich
constructor APIs (e.g., \texttt{PIL.Image.new} for JPEG/TIFF) that
make generator scripts compact, and (ii) the generated bytes are
opaque, so AFL++'s bit/byte mutators are appropriate downstream.

Structured text parsers exhibit the opposite properties. The
relevant edits at the syntactic level --- inserting a key, replacing
a literal with an interesting value, splicing two well-formed
sub-trees, applying a dictionary token --- are easier to express as
small \emph{data} (an opcode, a target offset, a replacement
literal) than as code. Furthermore, running LLM-emitted code at
runtime carries non-trivial safety risk (network calls, file I/O,
unbounded loops) that a schema-validated data structure avoids.
\S\ref{sec:recipe} develops this contrast in detail.

\subsection{Open design questions for structured text}
\label{sec:bg:openq}

Within the off-hot-path paradigm, three design questions remain
open for structured text parsers and motivate the rest of the
paper:

\begin{itemize}
  \item \textbf{Q1: Intermediate representation.} Code-as-generator
  or data-as-recipe? (\S\ref{sec:recipe})
  \item \textbf{Q2: Proposal validation.} How does the controller
  decide whether a candidate proposal is worth a multi-hour main-loop
  commitment? (\S\ref{sec:plateau})
  \item \textbf{Q3: Static context source.} When the target's source
  is unavailable, where does the structured context come
  from? (\S\ref{sec:plateau:ghidra})
\end{itemize}

\section{FuzzPilot Design}
\label{sec:design}

\begin{figure}[p]
  \centering
  \resizebox{\textwidth}{!}{\input{figures/F1_architecture.tikz}}
  \caption{FuzzPilot end-to-end architecture. Five horizontal stripes
  separate \textbf{external tools} (Ghidra, optional model client), the
  off-hot-path \textbf{controller} (plateau detector, blackboard,
  agents, micro-campaign runner), persisted \textbf{artifacts}
  (\texttt{static\_context.json}, \texttt{proposal.json},
  \texttt{recipe.json}, \texttt{main\_recipes/}), the native
  \textbf{hot path} (AFL++ runner, corpus, custom mutator, target
  binary), and per-run \textbf{outputs} (\texttt{fuzzer\_stats},
  \texttt{coverage.csv}, \texttt{events.jsonl}, audit log). Only two
  edges cross stripes (drawn in
  \textcolor{red!70!black}{red}): telemetry up (Step~3)
  and recipe promotion down (Step~9). No
  per-mutation hot-path code ever touches the optional model client. Numbered edges
  \protect\circled{1}--\protect\circled{9} are walked through in
  \S\ref{sec:design} and \S\ref{sec:plateau}.}
  \label{fig:arch}
\end{figure}

\subsection{Architectural contract}
\label{sec:design:contract}

The architecture (Figure~\ref{fig:arch}) is organized around a
single invariant: \emph{no external reasoning call appears on any code path
executed by AFL++ during per-mutation work.} Model calls occur
only when (a) the plateau detector marks a window as stalled, or
(b) a micro-campaign has completed and a promotion decision is
needed. The mutator never holds an open connection to the optional
model client; it consumes recipes from a local store that contains only
validated data.

This contract is shared with G\textsuperscript{2}FUZZ; we adopt it
as the baseline. The remainder of this section describes how
FuzzPilot specializes the contract for the structured-text-parser
domain.

\subsection{Component overview}
\label{sec:design:components}

\textbf{Plateau detector.} A coverage-window monitor that flags a
plateau when fewer than $\theta_e$ new executions and fewer than
$\theta_p$ new paths accumulate over $W$ seconds. Hyperparameters are
pinned per target (\S\ref{sec:plateau}).
On a plateau the detector freezes a snapshot of the queue and writes
it to the blackboard.

\textbf{Blackboard.} A JSON document holding the plateau snapshot,
recent fuzzer\_stats, and the Ghidra-derived static context for
the target binary (\S\ref{sec:plateau:ghidra}). The blackboard is
the LLM's only view of the campaign state.

\textbf{Proposal workers.} A small set of role-specialized prompt templates
(coordinator, plateau-diagnosis, scheduler, cmp-hint, dictionary,
format, mutator, corpus, crash-triage). Each agent reads the
blackboard and proposes a recipe or a sub-recipe; the coordinator
assembles a final proposal. We do not claim the role split itself as
a novelty in this preprint; it is an engineering convenience and is
ablated jointly with the other control-path components.

\textbf{Recipe store.} A versioned, schema-validated collection of
recipes proposed, evaluated, and (in some cases) promoted. The
custom mutator reads from this store at recipe-load time only.

\textbf{Micro-campaign runner.} Launches an isolated $N$-second
AFL++ run with a candidate recipe installed, computes a reward over
its delta in edges and paths, and reports back. The micro-campaign
shares the target binary with the main run but uses a private
working directory and a snapshot of the corpus.

\textbf{Custom mutator.} A native AFL++ mutator that dispatches on
the seven recipe operations (\texttt{BitFlip},
\texttt{OverwriteRange}, \texttt{InsertToken}, \texttt{Arith},
\texttt{Splice}, \texttt{DeleteBlock}, \texttt{DictionaryOverwrite};
the full enumeration is in \S\ref{sec:recipe:ops}). The mutator
is the only on-the-hot-path component
that touches recipes; its cost profile is reported in
\S\ref{sec:eval} (RQ2, F5).

\section{Recipe Abstraction: Data, Not Code}
\label{sec:recipe}

\subsection{Recipe schema}
\label{sec:recipe:schema}

A recipe carries two layers of state that the implementation keeps
in lock-step. The high-level form,
\texttt{SeedMutationStrategy}
(\texttt{include/fuzzpilot/mutation/strategy.hpp:38--49}), is what
the agent emits and the controller serializes; it is lowered to a
compact binary form, \texttt{CompactRecipe}
(\texttt{mutators/fuzzpilot/recipe\_cache.hpp:45--58}), before the
custom mutator loads it. The two share the following core fields:

\begin{itemize}
  \item \texttt{id}: stable identifier; lets repeat recipes
  deduplicate across plateaus and across runs.
  \item \texttt{selector}: which subset of corpus elements the
  recipe applies to. A selector is a (\emph{mode}, \emph{key}) pair
  where \emph{mode} is one of \texttt{mode}~/ \texttt{seed\_id}~/
  \texttt{seed\_hash}~/ \texttt{family} and \emph{key} is the
  matching string.
  \item \texttt{priority} and \texttt{ttl\_sec}: scheduling weight
  and time-to-live for the recipe store.
  \item \texttt{operator\_weights} (in compact form:
  \texttt{weights}): a probability distribution over the fixed
  7-operation vocabulary (\S\ref{sec:recipe:ops}). The mutator
  samples \emph{one} op per call by these weights; the recipe is
  not an ordered program.
  \item \texttt{focus\_ranges} / \texttt{protect\_ranges}: half-open
  byte ranges $[a,b)$ where the mutator is preferred (respectively
  forbidden) to write. Derived from agent inference and from the
  Ghidra-extracted \texttt{cmp\_constants} and
  \texttt{branch\_constraints} (\S\ref{sec:plateau:ghidra}).
  \item \texttt{dictionary\_tokens} (compact: \texttt{tokens}):
  literal byte sequences that the \texttt{InsertToken} and
  \texttt{DictionaryOverwrite} ops draw from. Populated from
  Ghidra \texttt{magic\_tokens} and from agent-proposed tokens.
  \item \texttt{expected\_signal}: the agent's prediction of what
  kind of coverage gain this recipe should produce. Used by the
  audit log only; never consulted by the mutator.
  \item \texttt{fields} (compact only): per-field overrides for
  structured parsers, e.g.~``JSON object values'' or ``PNG IHDR
  block''.
\end{itemize}

Recipes that fail JSON-schema validation are discarded before they
reach the recipe store. The mutator never parses free-form model
output --- it only loads \texttt{CompactRecipe} entries from a
local binary cache.

\subsection{Operation vocabulary}
\label{sec:recipe:ops}

The op vocabulary is small (seven ops) and closed
(\path{mutators/fuzzpilot/recipe_cache.hpp:9--17}):

\begin{itemize}
  \item \texttt{BitFlip} --- single-bit flip at a chosen offset.
  \item \texttt{OverwriteRange} --- write random bytes into a
  contiguous range.
  \item \texttt{InsertToken} --- splice a token from
  \texttt{dictionary\_tokens} at a chosen offset.
  \item \texttt{Arith} --- arithmetic on a numeric field
  (add / subtract / multiply / divide).
  \item \texttt{Splice} --- replace a sub-range with bytes lifted
  from another queue element.
  \item \texttt{DeleteBlock} --- excise a contiguous range.
  \item \texttt{DictionaryOverwrite} --- overwrite bytes at a
  chosen offset with a dictionary token.
\end{itemize}

Each call to the mutator's fuzz function invokes
\texttt{recipe.choose\_operator(rng)} to sample one op by
\texttt{operator\_weights} and dispatches it through a 7-arm
switch (\texttt{mutators/fuzzpilot/\allowbreak mutator\_core.cpp:\allowbreak 292--318}).
There is no opcode interpreter and no nested control flow: the
mutator's work per mutation is bounded by the cost of the chosen
op, which is the same order of magnitude as AFL++'s native havoc
operators. This keeps the throughput cost of recipe dispatch within
the budget we report in \S\ref{sec:eval:rq2}.

\subsection{Example recipe (illustrative)}
\label{sec:recipe:example}

Listing~\ref{lst:recipe} shows a recipe an agent might emit after
inspecting a JSON parser's Ghidra context (an opening brace
\texttt{0x7b} expected at offset 0; \texttt{strcpy} call site
detected at offset \texttt{0x108}). The mutator loads the recipe
from cache and, on each call, samples an op by
\texttt{operator\_weights} and applies it inside
\texttt{focus\_ranges}, drawing tokens from \texttt{dictionary\_tokens}
when an \texttt{InsertToken} or \texttt{DictionaryOverwrite} op is
chosen. The agent never re-enters the loop during mutation.

\begin{figure}[t]
\begin{lstlisting}[language=json,caption={Example recipe (cJSON, focus on object/array boundary tokens).},label={lst:recipe}]
{
  "id": "p1_e1b_r03_plateau_07",
  "selector": {"mode": "seed_hash", "key": "9c4f...a7b1"},
  "priority": 3,
  "ttl_sec": 1800,
  "operator_weights": {
    "InsertToken":         0.35,
    "DictionaryOverwrite": 0.25,
    "Splice":              0.15,
    "OverwriteRange":      0.10,
    "BitFlip":             0.10,
    "Arith":               0.03,
    "DeleteBlock":         0.02
  },
  "focus_ranges":   [[0, 1], [42, 64]],
  "protect_ranges": [[16, 20]],
  "dictionary_tokens": ["{", "}", "[", "]", "\"", "true", "null"],
  "expected_signal": "exercise object/array nesting boundary"
}
\end{lstlisting}
\end{figure}

\subsection{Recipe lifecycle}
\label{sec:recipe:lifecycle}

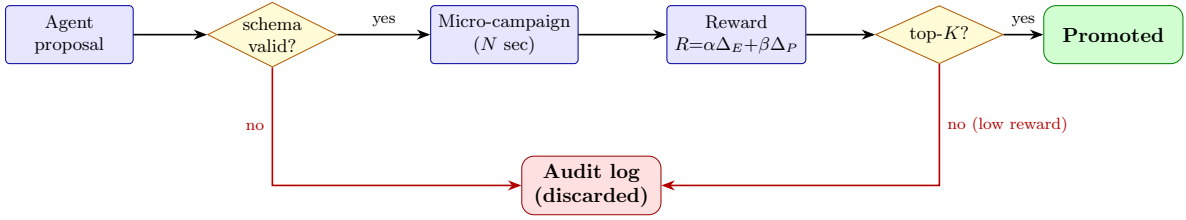
\begin{figure}[t]
  \centering
  \resizebox{0.95\textwidth}{!}{

\begin{tikzpicture}[
    font=\footnotesize,
    >=Stealth,
    state/.style={
        draw, rounded corners=2pt,
        minimum width=22mm, minimum height=10mm,
        align=center,
        fill=blue!10, draw=blue!60!black,
    },
    decision/.style={
        draw, diamond, aspect=2,
        inner sep=0pt,
        minimum width=22mm, minimum height=10mm,
        align=center,
        fill=yellow!20, draw=orange!70!black,
    },
    promoted/.style={
        draw, rounded corners=6pt,
        minimum width=24mm, minimum height=10mm,
        align=center,
        font=\small\bfseries,
        fill=green!18, draw=green!50!black,
    },
    discarded/.style={
        draw, rounded corners=6pt,
        minimum width=24mm, minimum height=10mm,
        align=center,
        font=\small\bfseries,
        fill=red!12, draw=red!60!black,
    },
    arr/.style={->, thick},
    noarr/.style={->, thick, red!70!black},
]

\node[state]    (proposal) at (0,    0)  {Agent\\proposal};
\node[decision] (schema)   at (3.5,  0)  {schema\\valid?};
\node[state]    (micro)    at (7.5,  0)  {Micro-campaign\\($N$~sec)};
\node[state]    (reward)   at (11.5, 0)  {Reward\\$R{=}\alpha\Delta_E{+}\beta\Delta_P$};
\node[decision] (rank)     at (15,   0)  {top-$K$?};
\node[promoted] (promoted) at (18,   0)  {Promoted};

\node[discarded] (audit) at (9, -2.6) {Audit log\\(discarded)};

\draw[arr] (proposal) -- (schema);
\draw[arr] (schema)   -- node[above, font=\scriptsize] {yes} (micro);
\draw[arr] (micro)    -- (reward);
\draw[arr] (reward)   -- (rank);
\draw[arr] (rank)     -- node[above, font=\scriptsize] {yes} (promoted);

\draw[noarr] (schema) |- node[pos=0.25, left, font=\scriptsize] {no} (audit);
\draw[noarr] (rank)   |- node[pos=0.25, right, font=\scriptsize] {no (low reward)} (audit);

\end{tikzpicture}}
  \caption{Recipe lifecycle. Each LLM proposal is first
  schema-validated, then evaluated in an $N$-second isolated
  micro-campaign, and only the top-$K$ by reward are promoted to
  the main run. Discarded recipes are kept in the audit log.}
  \label{fig:lifecycle}
\end{figure}

The lifecycle (Figure~\ref{fig:lifecycle}) is:
\textbf{(1) Proposal} from one or more agents; \textbf{(2) Schema
validation} (discard malformed proposals); \textbf{(3) Micro-campaign}
isolated $N$-second AFL++ run with the proposal installed;
\textbf{(4) Reward} computed over $\Delta$edges and $\Delta$paths;
\textbf{(5) Promotion} of top-$K$ proposals to the main recipe store,
others archived to the audit log.

\subsection{Why data, not code? Contrast with code-as-generator}
\label{sec:recipe:why}

G\textsuperscript{2}FUZZ's input-generator approach is the natural
intermediate for non-textual formats, but in the structured-text
setting data-as-recipe has three concrete advantages.

\textbf{Runtime safety.} Executing LLM-emitted Python at fuzz time
exposes the harness to a class of risks that schema-validated data
does not: import side effects, file I/O, network calls, and unbounded
loops are all reachable from a free-form generator. The recipe
dispatcher has no I/O capability beyond reading bytes from the
mutated buffer and writing bytes back.

\textbf{Cacheability and version control.} A recipe is a small
($\leq 2$~KB) JSON document with a stable schema; thousands of
recipes can be diffed, deduplicated, indexed by hash, and
re-evaluated without re-prompting the model. A generator script is
larger, depends on Python and library versions, and is awkward to
de-duplicate or diff.

\textbf{Schema-validatable in finite time.} The recipe schema is a
closed grammar; validation is a JSON-schema check. A generator
script is Turing-complete and cannot be statically validated for
behavior; G\textsuperscript{2}FUZZ relies on a syntactic-validator
plus a sandbox to compensate.

We do not claim data-as-recipe is universally superior; in the
non-textual setting it is plausibly less expressive. The claim is
narrower: for structured text parsers, data-as-recipe yields a safer
and more reusable intermediate than code-as-generator, and our
evaluation (\S\ref{sec:eval}) reports its concrete cost
(\texttt{fp-active} vs \texttt{fp-empty}).

\section{Plateau Detection, Ghidra Context, and Micro-Campaign}
\label{sec:plateau}

\subsection{Plateau detection}
\label{sec:plateau:detect}

The plateau detector inspects a sliding window of $W$ seconds and
flags a plateau when, within that window, both (a) fewer than
$\theta_e$ new executions and (b) fewer than $\theta_p$ new paths
have accumulated. The three values are pinned per target in the run
configuration (\path{experiments/targets/cjson/config.yaml}); for
cJSON we use $W=10\,\mathrm{s}$, $\theta_e=50$\,execs, $\theta_p=1$
path. In each of the 5 \texttt{full-agent} runs the detector fired
exactly one plateau cycle within the 14{,}400\,s budget (corroborated
by the \texttt{corpus\_snapshot} event count of 1 per run in
\texttt{events.jsonl}); cJSON is small enough that one plateau is
sufficient to use the productive-then-plateau window once. On a
flagged plateau the detector freezes a queue snapshot, writes it to
the blackboard, and signals the coordinator agent.

\subsection{Ghidra-derived static context}
\label{sec:plateau:ghidra}

\paragraph{Extraction pipeline.}
Once per target binary, FuzzPilot launches Ghidra
\texttt{analyzeHeadless} in an ephemeral project directory
(\texttt{<run\_dir>/ghidra\_project}, automatically deleted on
exit) and runs a custom post-script,
\texttt{FuzzPilotGhidraExtract.java}. The script writes
\texttt{static\_context.json} with the following schema:

\begin{itemize}
  \item \texttt{functions[]}: top-20 risky functions, each
  annotated with name, address, size, list of risky calls, and a
  danger score. The score weights direct calls to known unsafe
  primitives:
  $10\times \#\{\texttt{gets},\,\texttt{system},\,\texttt{popen}\}$
  $+ 5\times \#\{\texttt{strcpy},\,\texttt{strcat}\}$
  $+ 4\times \#\{\texttt{sprintf}\}$
  $+ 3\times \#\{\texttt{malloc},\,\texttt{free},\,\texttt{memcpy},\,\texttt{memmove}\}$
  $+ 2\times \#\{\texttt{printf}\}$.
  \item \texttt{magic\_tokens[]}: up to 150 candidate dictionary
  entries --- defined strings recovered from Ghidra's
  \texttt{getDefinedData()} pass and 4-byte ASCII constants, in both
  endiannesses.
  \item \texttt{cmp\_constants[]}: up to 100 scalar operands
  appearing in comparison instructions, useful for the
  \texttt{numeric\_interesting} op of the recipe schema.
  \item \texttt{branch\_constraints[]}: up to 150 tuples of the
  form (address, condition mnemonic, comparison mnemonic,
  operand 1, operand 2), recovered by a 4-instruction lookahead
  preceding each conditional jump.
  \item \texttt{decompiled\_logic\{\}}: the decompiled C source
  for the top-5 risky functions, truncated to 4~KB each, for
  prompt-readable context.
  \item \texttt{structs}: target-defined struct layouts when
  available.
\end{itemize}

\paragraph{Two consumers, one extraction.}
The same \texttt{static\_context.json} feeds two downstream
components without re-entering the model:

\begin{enumerate}
  \item \textbf{Agent blackboard.} The JSON is merged into the
  blackboard's \texttt{static\_analysis\_context} field and read
  by every agent prompt during plateau handling. The LLM sees the
  ranked risky functions and decompiled C as natural-language
  context, alongside the queue snapshot and recent telemetry.
  \item \textbf{Micro-campaign AFL dictionary.} The
  \texttt{magic\_tokens} and \texttt{cmp\_constants} are also
  converted into an AFL-format dictionary file
  (\texttt{static\_generated.dict}) that the micro-campaign passes
  to AFL++ via \texttt{-x}. This puts static-analysis-derived
  tokens directly in the hands of the in-loop mutator, with no
  per-mutation LLM call.
\end{enumerate}

This dual-consumer design is a deliberate choice: the LLM uses the
context for high-level proposal, while AFL++ uses it for low-level
token-level mutation, and both share a single, cacheable extraction.

\paragraph{Caching.}
The first run of a campaign emits
\texttt{<run\_dir>/static\_context.json} and, if no prior cache
exists for the binary, renames it to
\texttt{base\_intelligence.json}. Subsequent runs on the same
binary load the cached JSON rather than re-running Ghidra. A
4-KB decompilation cap and the top-20 / top-150 / top-100 limits
keep the artifact under 1~MB even for large targets.

\paragraph{Failure handling.}
If \texttt{analyzeHeadless} exits non-zero, FuzzPilot writes an
error envelope into \texttt{static\_context.json} rather than an
empty object. Downstream code sees the error explicitly, and the
audit log records the failure so a re-run can replay the same
condition. The plateau handler still proceeds (with an empty
static context); the LLM's proposals are noted as ``static
context unavailable.''

\paragraph{Why Ghidra (and an honest caveat).}
We choose Ghidra over LLVM IR or angr for three operational
reasons: (a) Ghidra accepts stripped binaries without source code
or rebuild flags, broadening applicability; (b) its decompiled C
output is closer to what a human reverse engineer would feed an
LLM, and our prompts are trained on that style; (c) Ghidra
headless is a stable production tool with predictable resource
use, suitable for unattended fuzz campaigns. The channel is
an \textbf{enabling component} of the architecture, not a
free-standing novelty: on our open-source targets, an LLVM-IR or
angr-based pipeline could in principle supply the same fields.
Ablation E2c (\S\ref{sec:eval:rq4-ablation}) quantifies the contribution
of the entire channel to coverage, jointly measuring both
consumers (agent blackboard and AFL dictionary); it does not
isolate Ghidra-versus-alternative-backend gains.

\paragraph{Channel yield per target (Table~\ref{tab:ghidra-yield}).}
The Ghidra channel's empirical value depends on how much
structurally meaningful token material the target binary actually
exposes. As a sanity check on this axis, we extract the .rodata
string-literal vocabulary from each target binary via
\texttt{objcopy --only-section=.rodata} followed by GNU
\texttt{strings} and a printable-ASCII filter
(\texttt{scripts/paper01/extract\_strings\_dict.py}). This is a
cheaper, library-free proxy for the broader Ghidra extraction
pipeline (which the script also wraps in
\texttt{extract\_ghidra\_dict.sh}); we use it here as the headline
``how big is the candidate dictionary'' number because it is
deterministic, language-agnostic, and easily reproducible without
a Ghidra install. Table~\ref{tab:ghidra-yield} reports the count
for each binary checked into the repository.
\emph{cJSON} yields only $40$ usable string literals --- mostly
JSON keywords ($\mathtt{null}$, $\mathtt{true}$, $\mathtt{false}$)
and format specifiers ($\mathtt{\%1.15g}$, $\mathtt{u\%04x}$) ---
which is one structural reason the LLM proposal layer produced a
null result on cJSON (\S\ref{sec:eval:rq3}): the static channel
simply has very little to work with. The venue targets yield
1{,}000$+$ to 6{,}000$+$ tokens (XML diagnostic strings, SQL
keywords, ASN.1 OID labels), which gives the LLM enough material
to propose target-specific recipes rather than rely on the
fixed-three-token DictionaryAgent fallback. We caution that
\emph{the raw token count is an upper bound on usable dictionary
candidates}: openssl in particular includes embedded test PEM
certificates and base64 blobs that survive the printable-ASCII
filter but are not semantically useful as fuzzer dictionary
entries; downstream consumers (AFL++ \texttt{-x} and the
FuzzPilot mutator) re-rank by per-token reward signal.

\begin{table}[H]
  \centering
  \footnotesize
  \setlength{\tabcolsep}{6pt}
  \begin{tabular}{@{}lrr Y{7.5cm}@{}}
    \toprule
    Target & Binary size & Unique tokens & Token examples (top by occurrence) \\
    \midrule
    cJSON              & $228$\,KB & $40$       & \texttt{null}, \texttt{true}, \texttt{false}, \texttt{\%1.15g}, \texttt{u\%04x} \\
    libxml2 (parser)   & $5.6$\,MB & $2{,}572$  & \texttt{"Start tag expected"}, \texttt{"PEReference forbidden"}, \texttt{"DOCTYPE improperly terminated"} \\
    sqlite3 (parser/VM)& $7.2$\,MB & $1{,}164$  & \texttt{"SQLite format 3"}, \texttt{":memory:"}, \texttt{"(join-\%u)"}, \texttt{"(subquery-\%u)"}, \texttt{unix-none} \\
    openssl X.509      & $12$\,MB  & $6{,}656$\,$^{\dagger}$ & ASN.1 OID labels + embedded test certificate blobs (mixed quality; see caveat below) \\
    \bottomrule
  \end{tabular}
  \caption{Per-target \texttt{.rodata} string-literal yield, on the
  binaries checked into \texttt{experiments/targets/}.
  $^{\dagger}$openssl's count is inflated by embedded test
  certificates and base64 blobs that survive the printable-ASCII
  filter but are not semantically useful as fuzzing tokens; the
  effective usable count is smaller. The cJSON row anchors the
  preprint null result (the static channel has only $\sim$40
  things to say to the LLM); the venue rows show that the channel
  has 30--160$\times$ more raw material to work with on richer
  binaries. Reproduce with
  \texttt{python3 scripts/paper01/extract\_strings\_dict.py <binary>}.}
  \label{tab:ghidra-yield}
\end{table}

\subsection{Micro-campaign protocol}
\label{sec:plateau:micro}

Each plateau cycle launches a fixed bundle of $K_{\text{cand}}$
candidate recipes (default $K_{\text{cand}}=4$, set in
\texttt{micro\_campaign.num\_candidates}), one per intervention type
(\texttt{default}, \texttt{dictionary}, \texttt{seed\_focus},
\texttt{per\_seed\_recipe}). Each candidate is evaluated as follows:

\begin{enumerate}
  \item Snapshot the current corpus and queue state to a private
  working directory.
  \item Launch AFL++ for $N$ seconds (default $N=20$, set in
  \texttt{micro\_campaign.budget\_sec}) with the candidate recipe
  installed in the recipe store. The micro-campaign uses an
  edge-coverage map disjoint from the main run.
  \item Compute reward
  \begin{equation*}
    R \;=\; \alpha\,\Delta_{\text{edges}} \;+\;
            \gamma\,\Delta_{\text{crashes}} \;+\;
            \delta_h\,h \;-\; \delta_m\,m,
  \end{equation*}
  where $h$ and $m$ are recipe-hit and recipe-miss counts; if the
  AFL bitmap is unavailable, $\beta\,\Delta_{\text{paths}}$
  substitutes for $\alpha\,\Delta_{\text{edges}}$. The weights are
  pinned in \texttt{src/micro/evaluator.cpp}: $\alpha{=}1.0$,
  $\gamma{=}10$, $\delta_h{=}10^{-3}$, $\delta_m{=}5{\times}10^{-4}$,
  $\beta{=}0.5$ (fallback only).
  \item Rank candidates by $R$ and promote the top scorer to the
  main recipe store \emph{only if} $R>0$; otherwise emit
  \path{winner_decided:status=no_significance} and
  \path{promotion_skipped:reason=no_successful_micro_campaign},
  leaving the main loop on the controller's default rule recipe.
\end{enumerate}

The micro-campaign trades a small amount of main-run budget
(typically $N$~seconds at the start of each plateau-triggered cycle,
$\leq 5\%$ of total wall-clock) for empirical evidence that a
proposal helps. Without the gate, every model proposal would race
into the main loop; bad proposals would degrade the campaign for an
indeterminate time before being noticed via aggregate metrics.
Ablation E2b (\texttt{no-mutator}, controller + LLM but AFL++'s default
mutator instead of the recipe-guided one) and E2c (\texttt{no-static-analysis},
LLM without Ghidra binary context) in \S\ref{sec:eval} quantify the
cost of trusting the model directly or removing the static channel.

\subsection{Determinism and reproducibility}
\label{sec:plateau:repro}

We pin the model temperature to $0.0$ and report per-run
schema-validity and fallback rates instead of relying on identical
model output across runs. Each step in the chain from blackboard
state through proposal, recipe, reward, and promotion is written to
an append-only audit log, indexed by
proposal hash. The full decision trail of any run is reproducible
end-to-end from the audit log plus the pinned corpus snapshot, even
when the underlying model is itself replaced.

\section{Preliminary Evaluation}
\label{sec:eval}
\subsection{Setup}
\label{sec:eval:setup}

\paragraph{Hardware and OS.}
All AFL++ campaigns and microbenchmarks run on a single 4-core Intel
Xeon E5-2699 v4 @ 2.20 GHz virtual machine with 8 GB RAM,
Ubuntu 24.04 LTS, kernel 6.8.0-48-generic, on a dedicated 80 GB data
volume. \texttt{/proc/sys/kernel/core\_pattern} is set to \texttt{core}
(default on this host is an apport pipe that AFL++ refuses to run
under).

\paragraph{Toolchain.}
AFL++ v4.21c built from source (pinned commit; instrumentation via
\texttt{afl-\allowbreak clang-\allowbreak fast}). Ghidra 11.1.2\_PUBLIC headless (OpenJDK 17).
FuzzPilot experiment runs at commit \texttt{85223d8} on branch
\texttt{main} (full SHA in \S\ref{sec:repro}), recorded in each
run's \texttt{run\_metadata.json}.
The manuscript itself may include later text-only edits; the
run-time code and raw artifacts are pinned by the per-run metadata.
Python deps (pandas 2.1.4, matplotlib 3.6.3, PyYAML 6.0.1) from
Ubuntu packages.

\paragraph{Target.}
The sole evaluated target is cJSON
(\texttt{experiments/targets/cjson/cjson\_fuzzer}, ELF 64-bit,
\texttt{afl-clang-fast} instrumented, \texttt{AFL\_MAP\_SIZE=65536}),
seeded from 26 hand-curated JSON files covering objects, arrays,
nested structures, Unicode escapes, numeric edge cases, and a few
intentionally malformed variants. The repository also contains a
libpng harness (\texttt{libpng\_fuzzer},
\texttt{AFL\_MAP\_SIZE=131072}, 21 seed PNGs) but its 4-hour
campaigns are not part of this preprint and are deferred to the
venue version.

\paragraph{Modes and repeats.}
Five modes share a single AFL++ binary and target build, differing
only in the controller's \texttt{--ablation} switch:
\texttt{baseline-afl} (no FuzzPilot controller, no model, no
mutator), \texttt{full-agent} (all subsystems enabled),
\texttt{rule-only} (controller + native mutator, LLM disabled),
\texttt{no-mutator} (controller + LLM, AFL++ default mutator),
\texttt{no-static-analysis} (controller + LLM + mutator, Ghidra
context muted). Repeats per cell: 5 for \texttt{baseline-afl} and
\texttt{full-agent}, 3 for the three ablation rows.

\paragraph{Budget.}
Each AFL++ run has \texttt{main\_budget\_sec = 14{,}400}
(4 h wall-clock). The acceptance gates in
\path{experiments/manifests/paper01_preprint.yaml} require
\texttt{run\_time} $\geq 14{,}000$\,s, \texttt{execs\_done}
$\geq 10^6$, at least 200 \texttt{coverage.csv} rows, all required
artifacts present (\texttt{fuzzer\_stats}, \texttt{coverage.csv},
\texttt{events.jsonl}, \texttt{run\_metadata.json},
\texttt{report.md}), and \texttt{status} equal to
\texttt{completed}.

\paragraph{Model.}
The reference model for \texttt{full-agent} runs is
\texttt{deepseek-chat} at temperature 0 via the OpenAI-compatible
endpoint \url{https://api.deepseek.com/chat/completions}. A
secondary configuration uses
\nimmodeid{} via NVIDIA NIM
(\url{https://integrate.api.nvidia.com}); the configuration switch is
documented in the reproducibility appendix. The microbench (E3,
\S\ref{sec:eval:rq2}) does not invoke any LLM.

\paragraph{Parallelism and isolation.}
AFL campaigns run 4-way parallel (one worker per core); microbench
runs (E3) are run serially with \emph{all AFL workers suspended via
\texttt{SIGSTOP}} for the duration of the microbench
($\approx 30$~s; resumed via \texttt{SIGCONT}). Empirically, running
the microbench either concurrently with AFL (CPU contention) or
4-way parallel against itself inflates per-rep variance by 2--3$\times$
and inverts the paper's central RQ2 ratio (\S\ref{sec:eval:rq2}); the
isolation methodology is therefore a prerequisite for sound microbench
numbers on a 4-core host. The
$\approx 30$~s SIGSTOP window consumes $< 0.2\%$ of each AFL run's
14{,}400~s budget and is logged in
\path{results/paper01_ai_recipe_mutation/runs/_logs/}.

\begin{table}[H]
  \centering
  \footnotesize
  \setlength{\tabcolsep}{6pt}
  \begin{tabular}{@{}l l@{}}
    \toprule
    \textbf{Component} & \textbf{Pinned value} \\
    \midrule
    Host CPU                 & 4-core Intel Xeon E5-2699 v4 @ 2.20 GHz \\
    Host RAM                 & 8 GB \\
    OS / kernel              & Ubuntu 24.04 LTS / Linux 6.8.0-48-generic \\
    Data volume              & 80 GB ext4 (\texttt{/dev/vdb1} on \texttt{/www}) \\
    \texttt{core\_pattern}   & \texttt{core} (default \texttt{|/usr/share/apport/...} overridden) \\
    AFL++                    & v4.21c, built from source, \path{afl-clang-fast} instrumentation \\
    Ghidra                   & 11.1.2\_PUBLIC headless, OpenJDK 17 \\
    FuzzPilot run commit     & \texttt{85223d8} (full SHA in \S\ref{sec:repro}) \\
    \texttt{AFL\_MAP\_SIZE}  & 65536 (cJSON) \\
    Python deps              & pandas 2.1.4, matplotlib 3.6.3, PyYAML 6.0.1 (Ubuntu pkgs) \\
    Reference LLM            & \texttt{deepseek-chat}, temperature 0, \texttt{api.deepseek.com} \\
    Fallback LLM             & \nimmodeid{} via NVIDIA NIM \\
    Per-run budget           & 14{,}400 s (4 h) wall-clock \\
    Repetitions              & 5 (\texttt{baseline-afl}, \texttt{full-agent}) / 3 (ablations) \\
    \bottomrule
  \end{tabular}
  \caption{Pinned evaluation setup. All numbers in this paper come
  from this single configuration. The reproducibility appendix
  records the run-time commit, target binary SHA-256, seed-corpus
  tarball checksum, and raw run-artifact paths from the fuzz cloud
  server.}
  \label{tab:setup}
\end{table}

\subsection{RQ1 --- Plateau recovery on cJSON (E1a vs E1b)}
\label{sec:eval:rq1}

\begin{figure}[t]
  \centering

\begin{tikzpicture}
\begin{axis}[
    width=0.92\textwidth,
    height=6.3cm,
    xmin=0, xmax=14400,
    ymin=263.5, ymax=271,
    xlabel={Wall-clock (seconds)},
    ylabel={Edges discovered},
    xlabel style={font=\small},
    ylabel style={font=\small},
    xticklabel style={font=\small},
    yticklabel style={font=\small},
    scaled x ticks=false,
    xtick={0,2000,4000,6000,8000,10000,12000,14400},
    xticklabels={0, 2\,000, 4\,000, 6\,000, 8\,000, 10\,000, 12\,000, 14\,400},
    ytick={264,266,268,269,270},
    yticklabels={264,266,268,269,270},
    ymajorgrids=true,
    xmajorgrids=true,
    grid style={dashed, gray!25, line width=0.4pt},
    axis lines=left,
    legend pos=south east,
    legend style={
        font=\small,
        fill=white, draw=gray!50,
        nodes={inner sep=2pt},
    },
    title={\small cJSON edge discovery after the initial ramp},
    title style={yshift=2pt},
    clip=true,
]

\draw[red!50!black, dashed, line width=0.8pt]
    (axis cs:0, 269) -- (axis cs:14400, 269);
\node[anchor=south east, font=\scriptsize, red!50!black]
    at (axis cs:14400, 269) {269-edge ceiling};

\addplot[
    color=blue!75!black,
    line width=1.2pt,
    mark=none,
] coordinates {
    (60,266) (120,267) (240,267) (360,267)
    (600,267) (840,268) (1080,268) (1320,268)
    (1680,268) (2040,268) (2400,268) (2760,268)
    (3120,268) (3480,268) (3840,268) (4200,269)
    (4560,269) (5280,269) (6000,269) (7000,269)
    (8000,269) (9000,269) (10000,269) (11000,269)
    (12000,269) (13000,269) (14000,269) (14400,269)
};
\addlegendentry{\texttt{baseline-afl} median (5 reps)}

\addplot[
    color=orange!85!black,
    line width=1.4pt,
    mark=none,
    dash pattern=on 5pt off 2pt,
] coordinates {
    (60,265) (120,266) (240,267) (600,267) (960,268)
    (1320,268) (1680,268) (2040,268) (2400,268)
    (2760,268) (3120,268) (3480,268) (3840,268)
    (4200,268) (4560,268) (4920,268) (5280,268)
    (5640,268) (6000,268) (6360,268) (6720,268)
    (7080,268) (7440,268) (7800,268) (8160,268)
    (8520,268) (8880,268) (9240,269) (9600,269)
    (10000,269) (11000,269) (12000,269) (13000,269)
    (14000,269) (14400,269)
};
\addlegendentry{\texttt{full-agent} median (5 reps)}

\draw[blue!70!black, dashed, line width=0.7pt]
    (axis cs:11911, 264) -- (axis cs:11911, 268.5);
\node[anchor=south, font=\scriptsize, blue!70!black, rotate=90]
    at (axis cs:11911, 266.6) {E1a median};

\draw[orange!85!black, dashed, line width=0.7pt]
    (axis cs:13059, 264) -- (axis cs:13059, 268.5);
\node[anchor=south, font=\scriptsize, orange!85!black, rotate=90]
    at (axis cs:13059, 266.6) {E1b median};

\end{axis}
\end{tikzpicture}
  \caption{Edge-discovery trajectory on cJSON for
  \texttt{baseline-afl} (E1a, $N{=}5$) and \texttt{full-agent}
  (E1b, $N{=}5$), zoomed into the 264--270 edge band after the
  initial ramp. Both modes converge to the same 269-edge ceiling;
  the difference is timing, not final coverage. The dashed verticals
  mark the per-mode median \texttt{last\_find}
  ($11{,}911$\,s vs $13{,}059$\,s), showing that
  \texttt{full-agent}'s productive phase extends closer to the budget
  end.}
  \label{fig:coverage}
\end{figure}
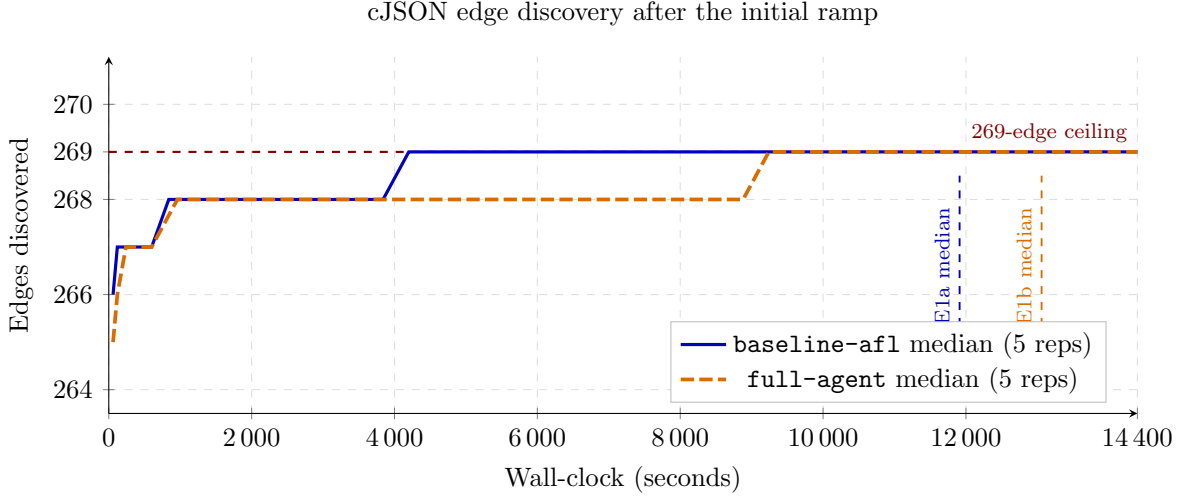

\paragraph{Edge ceiling on cJSON is a target-side property.}
Both modes saturate at the same 269-edge ceiling
(\texttt{edges\_found} field of AFL++'s \texttt{fuzzer\_stats},
also corroborated by \texttt{bitmap\_cvg} converging to
$23.21\%$ in all 10 runs) under the
14{,}400\,s budget: \texttt{baseline-afl} reaches 269 in 5 / 5
runs; \texttt{full-agent} reaches 269 in 4 / 5 runs and 266 in the
remaining one (r02). Figure~\ref{fig:coverage} plots both
median curves on the same axes; they converge to the dashed
ceiling line and do not exceed it in any mode. This is a property
of the cJSON harness, not of the fuzzer: cJSON exposes a limited
reachable-edge space that AFL++ alone exhausts well before the
budget, leaving no headroom for plateau-recovery improvement at
the \emph{ceiling} level. We therefore use cJSON for RQ2
(throughput parity) and RQ3 (LLM quality), and defer ceiling-level
RQ1 evidence to larger targets in the venue paper.

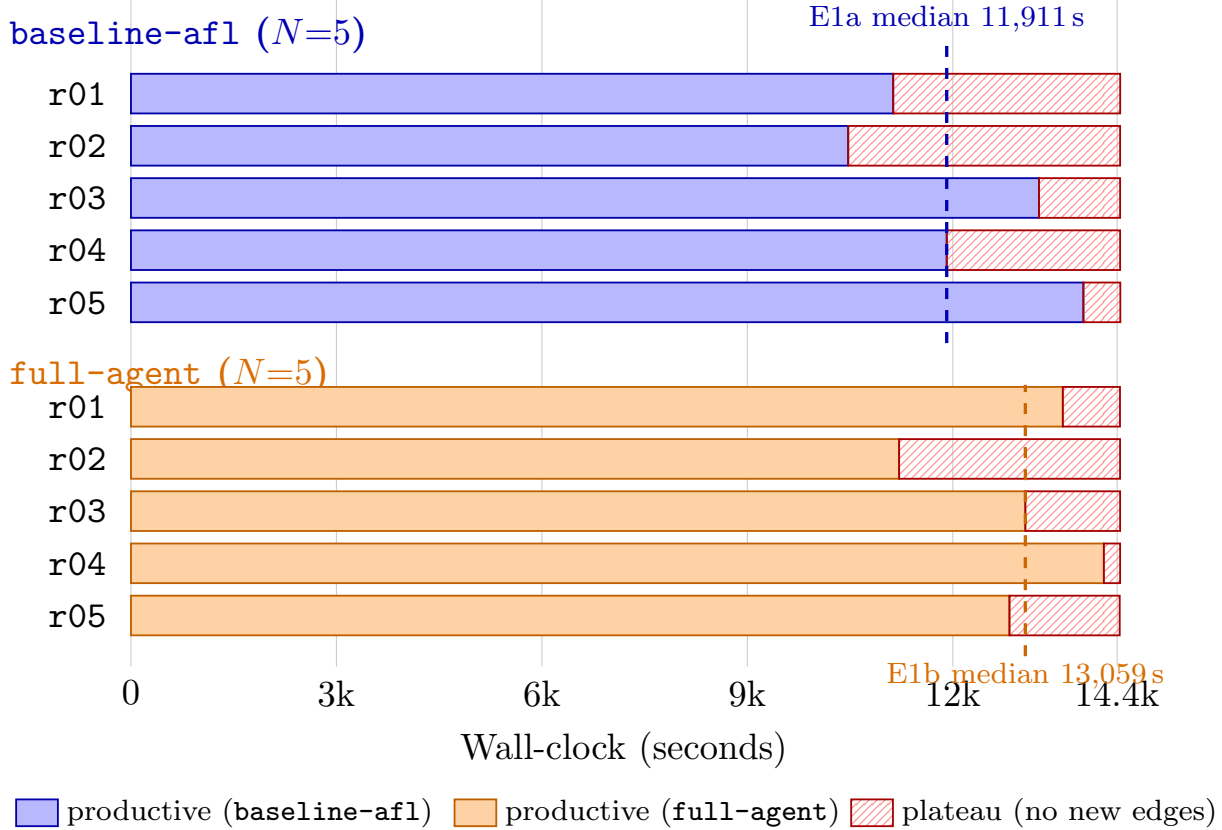
\begin{figure}[t]
  \centering
  \resizebox{\textwidth}{!}{
%
%

\begin{tikzpicture}[
    font=\small,
    >=Stealth,
    e1abar/.style={fill=blue!28, draw=blue!65!black, line width=0.5pt},
    e1bbar/.style={fill=orange!35, draw=orange!75!black, line width=0.5pt},
    plateaubar/.style={fill=red!15, pattern=north east lines, pattern color=red!45,
                        draw=red!65!black, line width=0.5pt},
    medianline/.style={dashed, line width=0.9pt},
]

\def\xscale{0.00065}
\def\yspacing{0.50}
\def\barH{0.38}
\def\labW{1.3}
\def\xrightedge{\labW + 14400 * \xscale}

\foreach \t/\lab in {0/0, 3000/3k, 6000/6k, 9000/9k, 12000/12k, 14400/14.4k} {
    \pgfmathsetmacro{\xx}{\labW + \t * \xscale}
    \draw[gray!40, line width=0.3pt] (\xx, {-1.6*\yspacing}) -- (\xx, {11.2*\yspacing});
    \node[anchor=north, font=\small] at (\xx, {-1.6*\yspacing}) {\lab};
}
\node[font=\small] at ({\labW + 7200*\xscale}, {-3.2*\yspacing})
    {Wall-clock (seconds)};

\node[anchor=west, font=\small\bfseries, blue!70!black]
    at (0, {10.5*\yspacing}) {\texttt{baseline-afl} ($N{=}5$)};
\node[anchor=west, font=\small\bfseries, orange!85!black]
    at (0, {4.0*\yspacing}) {\texttt{full-agent} ($N{=}5$)};

\newcommand{\drawrun}[5]{%
    \node[anchor=east, font=\small\ttfamily] at ({\labW - 0.10}, {#2*\yspacing + \barH/2}) {#1};
    \pgfmathsetmacro{\xlf}{\labW + #3 * \xscale}
    \pgfmathsetmacro{\xrt}{\labW + #4 * \xscale}
    \draw[#5] ({\labW}, {#2*\yspacing}) rectangle ({\xlf}, {#2*\yspacing + \barH});
    \draw[plateaubar] ({\xlf}, {#2*\yspacing}) rectangle ({\xrt}, {#2*\yspacing + \barH});
}

\drawrun{r01}{9}{11130}{14444}{e1abar}
\drawrun{r02}{8}{10473}{14443}{e1abar}
\drawrun{r03}{7}{13258}{14443}{e1abar}
\drawrun{r04}{6}{11911}{14443}{e1abar}
\drawrun{r05}{5}{13907}{14446}{e1abar}

\drawrun{r01}{3}{13606}{14441}{e1bbar}
\drawrun{r02}{2}{11216}{14442}{e1bbar}
\drawrun{r03}{1}{13059}{14443}{e1bbar}
\drawrun{r04}{0}{14205}{14443}{e1bbar}
\drawrun{r05}{-1}{12827}{14437}{e1bbar}

\pgfmathsetmacro{\xmA}{\labW + 11911 * \xscale}
\pgfmathsetmacro{\xmB}{\labW + 13059 * \xscale}
\draw[medianline, blue!70!black]
    ({\xmA}, {4.6*\yspacing}) -- ({\xmA}, {10.3*\yspacing});
\draw[medianline, orange!80!black]
    ({\xmB}, {-1.4*\yspacing}) -- ({\xmB}, {3.8*\yspacing});

\node[anchor=south, font=\scriptsize, blue!70!black]
    at ({\xmA}, {10.3*\yspacing})
    {E1a median 11{,}911\,s};
\node[anchor=south, font=\scriptsize, orange!85!black]
    at ({\xmB}, {-2.3*\yspacing})
    {E1b median 13{,}059\,s};

\def\legendY{-4.4*\yspacing}
\node[e1abar, minimum width=0.40cm, minimum height=0.18cm]
    at (0.4, \legendY) {};
\node[anchor=west, font=\scriptsize] at (0.55, \legendY)
    {productive (\texttt{baseline-afl})};
\node[e1bbar, minimum width=0.40cm, minimum height=0.18cm]
    at (4.6, \legendY) {};
\node[anchor=west, font=\scriptsize] at (4.75, \legendY)
    {productive (\texttt{full-agent})};
\node[plateaubar, minimum width=0.40cm, minimum height=0.18cm]
    at (8.4, \legendY) {};
\node[anchor=west, font=\scriptsize] at (8.55, \legendY)
    {plateau (no new edges)};

\end{tikzpicture}}
  \caption{Per-run plateau-onset timeline for E1a (top, blue) and
  E1b (bottom, orange). Solid bars show productive fuzzing through
  each run's \texttt{last\_find}; hatched bars show the trailing
  plateau to the budget end. Dashed lines mark the per-mode median
  \texttt{last\_find} ($11{,}911$\,s vs $13{,}059$\,s).}
  \label{fig:plateau-timeline}
\end{figure}

\paragraph{Plateau-duration delta is the real RQ1 signal.}
Even when both modes saturate to the same ceiling, the moment of
saturation differs systematically. Table~\ref{tab:plateau-delta}
and Figure~\ref{fig:plateau-timeline} report the median
\texttt{last\_find} (wall-clock at which the last new edge was
discovered, i.e.~end of the productive phase) and the resulting
\emph{plateau} duration (gap between \texttt{last\_find} and the
14{,}400\,s budget end). The full-agent group extends the
productive phase by $1{,}148$\,s ($\approx 19$\,min) at the
median, corresponding to a descriptive $45.3\%$ reduction in
plateau duration ($2{,}532 \to 1{,}384$\,s). \emph{This delta is
descriptive only and not statistically significant}: with $N{=}5$
per arm the per-run plateau distributions $[539, 1{,}185, 2{,}532,
3{,}314, 3{,}970]$\,s and $[238, 835, 1{,}384, 1{,}610, 3{,}226]$\,s
overlap substantially, and a two-sided Mann--Whitney $U$ test
(following~\cite{klees2018evaluating,arcuri2011practical}) gives
$U{=}8$, $p{=}0.42$; the Vargha--Delaney
$A_{12}{=}0.68$~\cite{vargha2000critique} indicates a moderate
effect in the direction the median suggests, but the bootstrap
95\% CIs for the two medians ($[539, 3{,}970]$\,s for baseline,
$[238, 3{,}226]$\,s for full-agent) almost entirely overlap. We
report the descriptive $-45.3\%$ as a point estimate that requires
validation at $N{\geq}20$ on non-saturated targets before it can be
claimed as a reproducible effect.

\paragraph{Attribution caveat: controller vs mutator.}
We initially attributed the plateau delta to the recipe-guided
mutator framework. However, the ablation experiments in
\S\ref{sec:eval:rq4-ablation} reveal a more complex picture: the
\path{no-mutator} ablation (E2b, which \emph{removes} the
recipe-guided mutator but keeps the controller) shows a
\emph{shorter} median plateau ($930$\,s) than \texttt{rule-only}
(E2a, $1{,}165$\,s) or \texttt{full-agent} (E1b, $1{,}384$\,s). If
the mutator were the sole driver, removing it should worsen
performance, not improve it. The most data-consistent interpretation
is that the \textbf{controller's plateau-detection and
corpus-reorganization machinery} is the primary driver, with the
recipe-guided mutator adding latency (slower \texttt{cycles\_done})
without corresponding plateau benefit on this saturated target. A
\path{controller-only} ablation (plateau detector + corpus snapshot
only, no mutator, no LLM, no Ghidra) is required to test this
hypothesis but is not in the preprint matrix; it is deferred to the
venue version (\S\ref{sec:limits}). Until that ablation is run, the
attribution remains a hypothesis rather than a measured effect.
E1b vs E1a adds the LLM and micro-campaign \emph{infrastructure
cost}, which Table~\ref{tab:plateau-delta} bounds at well under
the observed plateau gain. \emph{However, as
\S\ref{sec:eval:rq4-ablation} discusses, the ablation ordering
(\path{no-mutator} median plateau $930$\,s $<$ \texttt{rule-only}
$1{,}165$\,s $<$ \texttt{full-agent} $1{,}384$\,s) contradicts the
naive ``mutator-framework drives the gain'' attribution and points
toward the controller's plateau detector and corpus-snapshot
reorganization as a co-driver; the attribution remains a
descriptive hypothesis at this $N$.}

\begin{table}[t]
  \centering
  \footnotesize
  \setlength{\tabcolsep}{5pt}
  \begin{tabular}{@{}lrrrrr@{}}
    \toprule
    Mode & \makecell[r]{median\\ \texttt{last\_find}\,(s)} & \makecell[r]{median\\ plateau\,(s)} & \makecell[r]{95\% CI\\ plateau} & \makecell[r]{$\Delta$ plateau\\ (descriptive)} & \makecell[r]{Mann--Whitney $U$\\ ($p$, $A_{12}$)} \\
    \midrule
    \texttt{baseline-afl} (E1a, $N{=}5$) & 11{,}911 & 2{,}532 & $[539, 3{,}970]$ & --- & --- \\
    \texttt{full-agent}   (E1b, $N{=}5$) & 13{,}059 & 1{,}384 & $[238, 3{,}226]$ & $-1{,}148$ ($-45.3\%$) & 8 ($0.42$, $0.68$) \\
    \bottomrule
  \end{tabular}
  \caption{Plateau-duration delta on cJSON. Medians and deltas are
  descriptive: the distributions overlap, the Mann--Whitney result is
  not significant at $N{=}5$, and the bootstrap median CIs overlap.
  Figure~\ref{fig:plateau-timeline} shows the raw per-run values.}
  \label{tab:plateau-delta}
\end{table}

\paragraph{Per-run timeline.}
Figure~\ref{fig:plateau-timeline} shows the full per-run picture.
Three full-agent runs (r01, r03, r04) push past the
\emph{baseline median} ($11{,}911$\,s) and stay productive
significantly longer; one (r04) is productive until
$14{,}205$\,s. Two full-agent runs (r02, r05) plateau earlier
than the baseline best ($13{,}907$\,s); this is expected on a
saturated target, where which corpus subset a worker ends up in
dominates over agent quality. With $N{=}5$ reps the per-mode
signal is qualitative; the venue paper will repeat this on
Magma-suite targets at $N{\geq}10$ with formal Mann--Whitney $U$
and Vargha--Delaney $A_{12}$ reporting, following Klees
et~al.~\cite{klees2018evaluating}'s evaluation guidelines.

\paragraph{Statistical methodology disclosure.}
\label{sec:eval:rq1:stats}
For the descriptive comparisons above and in
\S\ref{sec:eval:rq4-ablation} we use the two-sided Mann--Whitney
$U$ test (\path{scipy.stats.mannwhitneyu}, exact for
$N{\leq}8$) for non-parametric difference in medians, the
Vargha--Delaney $A_{12}$~\cite{vargha2000critique} for effect
size, and the percentile bootstrap (10{,}000 resamples) for
median confidence intervals. The code that produces all reported
numbers (\path{scripts/paper01/review_stats.py}) is in the
repository and run from the same per-run \texttt{fuzzer\_stats}
files cited in the reproducibility appendix
(\S\ref{sec:repro}). Per
Klees et~al.~\cite{klees2018evaluating}, the canonical evaluation
guideline calls for $N{\geq}20$ repetitions; \emph{this preprint's
$N{=}5$ / $N{=}3$ does not meet that bar} and we therefore present
all comparisons as descriptive point estimates with their
non-parametric companions, rather than as significance claims.

\paragraph{Time-to-$N$-edges (Klees-style endpoint-free metric).}
Endpoint metrics like plateau duration depend only on
\texttt{last\_find} and ignore the shape of the discovery curve;
Klees et~al.~\cite{klees2018evaluating} \S6.1 caution against
relying on them. We therefore report
Table~\ref{tab:time-to-n}: the wall-clock at which each run
first reaches $N$ edges, for $N \in \{264, 268, 269\}$
(264 is one edge below the steady-state ramp, 268 is the
penultimate edge, 269 is the ceiling).

\begin{table}[t]
  \centering
  \footnotesize
  \setlength{\tabcolsep}{6pt}
  \begin{tabular}{@{}lrrr@{}}
    \toprule
    Mode & \makecell[r]{median time to\\ 264 edges (s)} & \makecell[r]{median time to\\ 268 edges (s)} & \makecell[r]{median time to\\ 269 edges (s)} \\
    \midrule
    \texttt{baseline-afl} ($N{=}5$)  & 60  & 660 & 2{,}524 \\
    \texttt{full-agent} ($N{=}5$)    & 120 & 662 & 6{,}702$^{\dagger}$ \\
    \texttt{rule-only} ($N{=}3$)     & 60  & 661 & 1{,}923 \\
    \texttt{no-mutator} ($N{=}3$)    & 60  & 601 & 1{,}804 \\
    \bottomrule
  \end{tabular}
  \caption{Median per-run wall-clock to reach $N$ edges.
  $^{\dagger}$One \texttt{full-agent} run (r02) never reaches 269
  within budget; the median is over the four runs that do
  $\{3{,}487, 4{,}329, 9{,}076, 14{,}245\}$\,s. All four ablation
  modes reach 264 and 268 in roughly the same time as
  \texttt{baseline-afl}; the gap opens at the 269 ceiling, where
  \texttt{baseline-afl} arrives \emph{earlier} on average than
  \texttt{full-agent} but then plateaus longer.}
  \label{tab:time-to-n}
\end{table}

The complementary endpoint-free reading is that the
\emph{ramp-up} (time to 264) is mode-independent; the
\emph{tail} (time to 269) is where modes diverge, and
\texttt{baseline-afl} actually reaches the ceiling
\emph{faster} at the median than \texttt{full-agent}.
\texttt{full-agent}'s advantage in plateau duration
(Table~\ref{tab:plateau-delta}) therefore comes from
discovering its final edge \emph{later in the budget}, not from
discovering the ceiling earlier. The two readings ---
``plateau is shorter'' and ``ceiling is reached later'' --- are
internally consistent: a 14{,}400\,s budget that ends with a new
edge at $14{,}205$\,s (full-agent r04) has a tiny plateau but a
very late ceiling-touch. A non-saturated target would let the
two metrics diverge in informative directions.

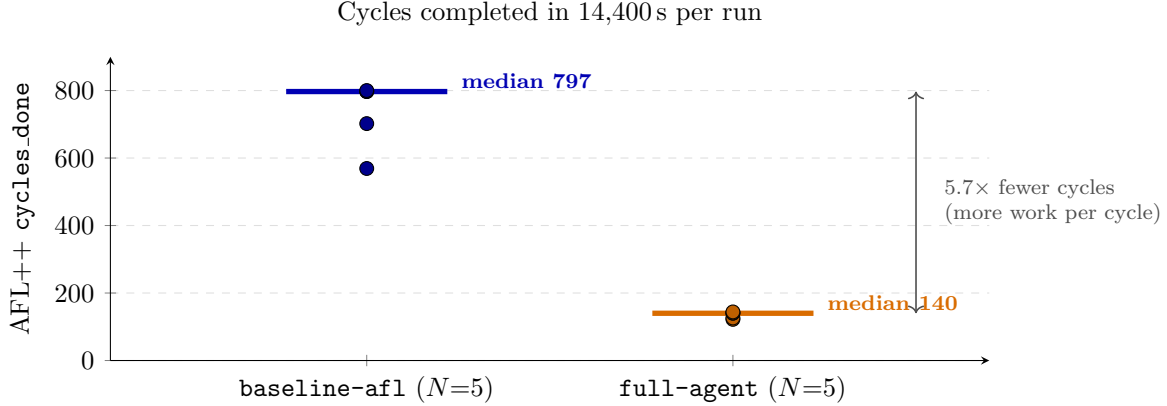
\begin{figure}[t]
  \centering
%

\begin{tikzpicture}
\begin{axis}[
    width=0.8\textwidth,
    height=5.6cm,
    xmin=0.3, xmax=2.7,
    ymin=0, ymax=900,
    ylabel={AFL++ \texttt{cycles\_done}},
    ylabel style={font=\small},
    xtick={1, 2},
    xticklabels={\texttt{baseline-afl} ($N{=}5$), \texttt{full-agent} ($N{=}5$)},
    xticklabel style={font=\small},
    yticklabel style={font=\small},
    ymajorgrids=true,
    grid style={dashed, gray!25, line width=0.4pt},
    axis lines=left,
    title={\small Cycles completed in 14{,}400\,s per run},
    title style={yshift=2pt},
    clip=false,
]
\addplot+[
    only marks,
    mark=*,
    mark size=2.6pt,
    mark options={fill=blue!55!black, draw=black, line width=0.4pt},
] coordinates {
    (1, 569) (1, 702) (1, 797) (1, 797) (1, 800)
};

\addplot+[
    only marks,
    mark=*,
    mark size=2.6pt,
    mark options={fill=orange!75!black, draw=black, line width=0.4pt},
] coordinates {
    (2, 122) (2, 126) (2, 140) (2, 141) (2, 144)
};

\draw[blue!70!black, line width=2pt]
    (axis cs:0.78, 797) -- (axis cs:1.22, 797);
\node[anchor=south west, font=\scriptsize\bfseries, blue!70!black,
      inner sep=1pt]
    at (axis cs:1.25, 797) {median 797};

\draw[orange!85!black, line width=2pt]
    (axis cs:1.78, 140) -- (axis cs:2.22, 140);
\node[anchor=south west, font=\scriptsize\bfseries, orange!85!black,
      inner sep=1pt]
    at (axis cs:2.25, 140) {median 140};

\draw[<->, gray!60!black, line width=0.6pt]
    (axis cs:2.5, 140) -- (axis cs:2.5, 797);
\node[anchor=west, font=\scriptsize, gray!60!black, align=left]
    at (axis cs:2.55, 470) {$5.7\times$ fewer cycles\\(more work per cycle)};

\end{axis}
\end{tikzpicture}
  \caption{\texttt{cycles\_done} per run across the two modes
  ($N{=}5$ each). Solid horizontal bars are per-mode medians.
  \texttt{full-agent} completes
  $\approx 5.7\times$ fewer corpus cycles in the same 14{,}400\,s
  budget, despite an \emph{equal-or-higher} total execution count
  (Figure~\ref{fig:throughput-dot}): the recipe-guided mutator
  spends more work per corpus item before recycling.}
  \label{fig:cycles}
\end{figure}

\paragraph{Mutation-rate redistribution (per-cycle vs per-execution).}
A second consequence of the recipe-guided mutator is visible in
\texttt{cycles\_done} (Figure~\ref{fig:cycles}). The
\texttt{baseline-afl} median is $797$ corpus cycles inside the
14{,}400\,s budget; the \texttt{full-agent} median is $140$, and
the two ablations that engage the same mutator
(\texttt{rule-only}: $119$; \texttt{full-agent}: $140$) sit in
that same $\approx 5\!-\!7\times$-fewer range, whereas
\path{no-mutator} ($671$) tracks much closer to
\texttt{baseline-afl}. The ratio is therefore attributable to the
recipe-guided mutator itself, not to any LLM-introduced step.
With a comparable total execution count (within $6\%$; RQ2,
Figure~\ref{fig:throughput-dot}), full-agent spends $\approx 5.7\times$
more executions on each pass through the queue before recycling to
the top --- exactly the mutation-rate redistribution the default
recipe encodes (\emph{focus dictionary-token insertions and overwrites
over the first $4\,$KB}, with fixed operator weights pinned at
\texttt{InsertToken}=0.35 / \texttt{DictionaryOverwrite}=0.25 /
\texttt{Splice}=0.15 / \texttt{OverwriteRange}=0.10 /
\texttt{BitFlip}=0.10 / \texttt{Arith}=0.03 /
\texttt{DeleteBlock}=0.02, matching Listing~\ref{lst:recipe}).
\emph{We caution that \texttt{cycles\_done} is not a cross-mode
comparator for ``progress''}: it is a derived quantity of the
form $\text{execs\_done} / \overline{\text{execs-per-cycle}}$,
where the denominator depends on per-input fuzzing intensity and
not on coverage progress. Two modes can run the same total
executions and reach the same edge ceiling with very different
\texttt{cycles\_done} values; the apparent ``5.7$\times$ less
work per cycle'' interpretation reduces to ``the recipe-guided
mutator runs more per-input mutation steps before popping the
queue head'', which is a definitional property of the recipe and
not a finding. Whether the concentration is empirically
beneficial is a target-specific question; on cJSON it correlates
with the longer productive phase
(Figure~\ref{fig:plateau-timeline}) but not with a larger final
edge set, because the target ceiling at 269 binds.

\paragraph{Per-run summary table (T1).}
Table~\ref{tab:t1} reports the 10 completed runs (5 baseline + 5
full-agent) at end-of-budget. Acceptance gates pass on all 10
(\texttt{run\_time} $\geq 14{,}000$\,s,
\texttt{execs\_done} $\geq 10^6$); zero saved crashes in either
mode (expected for cJSON's safety-tested code base).

\begin{table}[t]
  \centering
  \footnotesize
  \setlength{\tabcolsep}{4.5pt}
  \begin{tabular}{@{}llrrrrrrr@{}}
    \toprule
    Mode & Run & \texttt{run\_time} & \texttt{execs\_done} & exec/s & cycles & corpus & edges & \texttt{bitmap\_cvg} \\
    \midrule
    \multirow{6}{*}{\texttt{baseline-afl}}
       & r01 & 14{,}444 & 188{,}492{,}319 & 13{,}049 & 797 & 585 & 269 & 23.21\% \\
       & r02 & 14{,}443 & 187{,}158{,}013 & 12{,}958 & 800 & 607 & 269 & 23.21\% \\
       & r03 & 14{,}443 & 189{,}334{,}714 & 13{,}108 & 702 & 606 & 269 & 23.21\% \\
       & r04 & 14{,}443 & 185{,}251{,}854 & 12{,}826 & 569 & 649 & 269 & 23.21\% \\
       & r05 & 14{,}446 & 200{,}147{,}594 & 13{,}854 & 819 & 607 & 269 & 23.21\% \\
       & \emph{median} & 14{,}443 & 188{,}492{,}319 & 13{,}049 & 797 & 607 & 269 & 23.21\% \\
    \midrule
    \multirow{6}{*}{\texttt{full-agent}}
       & r01 & 14{,}441 & 199{,}520{,}607 & 13{,}816 & 122 & 598 & 269 & 23.21\% \\
       & r02 & 14{,}442 & 195{,}572{,}895 & 13{,}542 & 141 & 573 & \textbf{266} & 22.95\% \\
       & r03 & 14{,}443 & 195{,}567{,}575 & 13{,}540 & 144 & 601 & 269 & 23.21\% \\
       & r04 & 14{,}443 & 201{,}466{,}942 & 13{,}949 & 126 & 604 & 269 & 23.21\% \\
       & r05 & 14{,}437 & 225{,}180{,}431 & 15{,}597 & 140 & 600 & 269 & 23.21\% \\
       & \emph{median} & 14{,}442 & 199{,}520{,}607 & 13{,}816 & 140 & 600 & 269 & 23.21\% \\
    \bottomrule
  \end{tabular}
  \caption{T1: per-run end-of-budget summary for the 10 completed
  E1a + E1b cJSON runs. All runs satisfy the time and execution-count
  acceptance gates; RQ2 discusses the \texttt{exec/s} comparison.}
  \label{tab:t1}
\end{table}

\paragraph{Extended per-run statistics (T1-ext).}
Table~\ref{tab:t1-ext} reports the time-budget breakdown, target
stability, and plateau metrics for all 10 completed runs (5
baseline + 5 full-agent). The fuzz-time fraction
(\texttt{fuzz\_time}/\texttt{run\_time}) is uniformly $\geq 99.7\%$
in both modes, indicating that calibration, trim, and sync
overheads --- as well as the $\approx 30$\,s SIGSTOP windows from
the E3 microbench (\S\ref{sec:eval:rq2}) --- did not
noticeably perturb the AFL campaigns. Target stability (AFL++'s
coverage-bitmap stability check) is $\geq 99.25\%$ across all 10
runs, well within the standard ``stable target'' threshold of
90\%, so all measurements are comparable across modes.

\begin{table}[t]
  \centering
  \footnotesize
  \setlength{\tabcolsep}{6pt}
  \begin{tabular}{@{}llrrrrrr@{}}
    \toprule
    Mode & Run & \texttt{fuzz\_time} & \texttt{run\_time} & fuzz \% & stability & \texttt{last\_find}\,(s) & plateau\,(s) \\
    \midrule
    \multirow{6}{*}{\texttt{baseline-afl}}
       & r01 & 14{,}424 & 14{,}444 & 99.86\% & 99.26\% & 11{,}130 & 3{,}314 \\
       & r02 & 14{,}432 & 14{,}443 & 99.92\% & 99.26\% & 10{,}473 & 3{,}970 \\
       & r03 & 14{,}424 & 14{,}443 & 99.87\% & 99.26\% & 13{,}258 & 1{,}185 \\
       & r04 & 14{,}402 & 14{,}443 & 99.72\% & 99.26\% & 11{,}911 & 2{,}532 \\
       & r05 & 14{,}427 & 14{,}446 & 99.87\% & 99.26\% & 13{,}907 & \phantom{0}\phantom{,}539 \\
       & \emph{median} & 14{,}424 & 14{,}443 & 99.87\% & 99.26\% & \textbf{11{,}911} & \textbf{2{,}532} \\
    \midrule
    \multirow{6}{*}{\texttt{full-agent}}
       & r01 & 14{,}415 & 14{,}441 & 99.82\% & 99.26\% & 13{,}606 & \phantom{0}\phantom{,}835 \\
       & r02 & 14{,}421 & 14{,}442 & 99.85\% & 99.25\% & 11{,}216 & 3{,}226 \\
       & r03 & 14{,}425 & 14{,}443 & 99.88\% & 99.26\% & 13{,}059 & 1{,}384 \\
       & r04 & 14{,}417 & 14{,}443 & 99.82\% & 99.26\% & 14{,}205 & \phantom{0}\phantom{,}238 \\
       & r05 & 14{,}423 & 14{,}437 & 99.90\% & 99.26\% & 12{,}827 & 1{,}610 \\
       & \emph{median} & 14{,}421 & 14{,}442 & 99.85\% & 99.26\% & \textbf{13{,}059} & \textbf{1{,}384} \\
    \bottomrule
  \end{tabular}
  \caption{T1-ext: per-run fuzz-time fraction, target stability,
  \texttt{last\_find}, and plateau duration for the 10 completed
  E1a + E1b runs. The RQ1 text interprets the descriptive plateau
  delta.}
  \label{tab:t1-ext}
\end{table}

\paragraph{Cross-run throughput consistency.}
Within \texttt{baseline-afl}, the 5-run \texttt{execs\_per\_sec}
spread is $\{12{,}826, 12{,}958, 13{,}049, 13{,}108, 13{,}854\}$,
CV $= 3.1\%$ (median 13{,}049). Within \texttt{full-agent}, the
5-run spread is
$\{13{,}540, 13{,}542, 13{,}816, 13{,}949, 15{,}597\}$,
CV $= 6.1\%$ (median 13{,}816). Both groups satisfy the
\texttt{execs\_done} $\geq 10^6$ and \texttt{run\_time} $\geq 14{,}000$
acceptance gates with $> 10^3\times$ and $\sim 1.03\times$ margins
respectively. The end-to-end RQ2 throughput-parity ratio
(\S\ref{sec:eval:rq2}) is computed from these numbers.

\subsection{RQ2 --- Mutator throughput parity (F5 microbench)}
\label{sec:eval:rq2}

RQ2 has two pieces of evidence. End-to-end \texttt{execs\_per\_sec}
parity between \texttt{baseline-afl} (E1a) and \texttt{full-agent}
(E1b), measured by the runtime
\texttt{fuzzer\_stats:execs\_per\_sec} field over the 14{,}400\,s
budget, is the canonical RQ2 number; Table~\ref{tab:throughput-parity}
reports it. The microbench (E3) measures the dispatch cost of the
recipe-guided mutator alone (no AFL, no target); the rest of this
subsection covers the microbench.

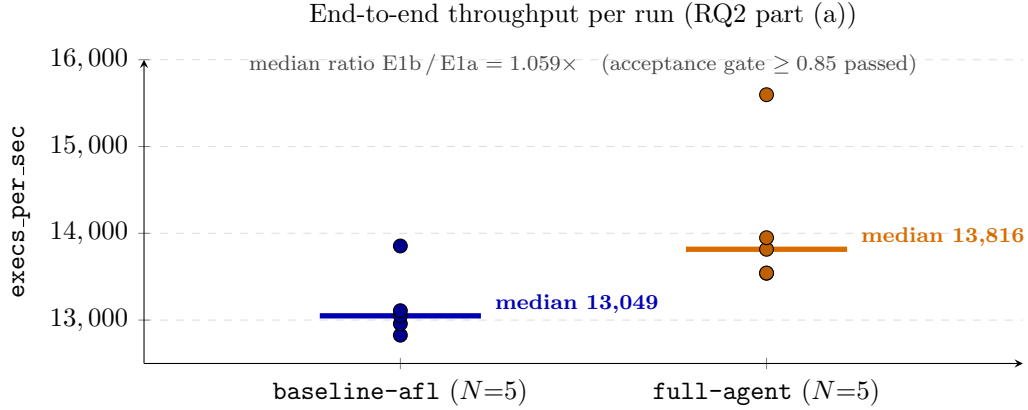
\begin{figure}[t]
  \centering
%

\begin{tikzpicture}
\begin{axis}[
    width=0.8\textwidth,
    height=5.6cm,
    xmin=0.3, xmax=2.7,
    ymin=12500, ymax=16000,
    ylabel={\texttt{execs\_per\_sec}},
    ylabel style={font=\small},
    xtick={1, 2},
    xticklabels={\texttt{baseline-afl} ($N{=}5$), \texttt{full-agent} ($N{=}5$)},
    xticklabel style={font=\small},
    yticklabel style={font=\small},
    ymajorgrids=true,
    grid style={dashed, gray!25, line width=0.4pt},
    axis lines=left,
    scaled y ticks=false,
    yticklabel={\pgfmathprintnumber[fixed,precision=0,1000 sep={,}]{\tick}},
    title={\small End-to-end throughput per run (RQ2 part (a))},
    title style={yshift=2pt},
    clip=false,
]
\addplot+[
    only marks,
    mark=*,
    mark size=2.6pt,
    mark options={fill=blue!55!black, draw=black, line width=0.4pt},
] coordinates {
    (1, 12826) (1, 12958) (1, 13049) (1, 13108) (1, 13854)
};

\addplot+[
    only marks,
    mark=*,
    mark size=2.6pt,
    mark options={fill=orange!75!black, draw=black, line width=0.4pt},
] coordinates {
    (2, 13540) (2, 13542) (2, 13816) (2, 13949) (2, 15597)
};

\draw[blue!70!black, line width=2pt]
    (axis cs:0.78, 13049) -- (axis cs:1.22, 13049);
\node[anchor=south west, font=\scriptsize\bfseries, blue!70!black,
      inner sep=1pt]
    at (axis cs:1.25, 13049) {median 13{,}049};

\draw[orange!85!black, line width=2pt]
    (axis cs:1.78, 13816) -- (axis cs:2.22, 13816);
\node[anchor=south west, font=\scriptsize\bfseries, orange!85!black,
      inner sep=1pt]
    at (axis cs:2.25, 13816) {median 13{,}816};

\node[anchor=south, font=\scriptsize, gray!55!black, align=center]
    at (axis cs:1.5, 15700)
    {median ratio E1b\,/\,E1a $= 1.059\times$\quad
     (acceptance gate $\geq 0.85$ passed)};

\end{axis}
\end{tikzpicture}
  \caption{End-to-end \texttt{execs\_per\_sec} per run for both
  modes ($N{=}5$ each); horizontal bars are per-mode medians.
  Even the slowest \texttt{full-agent} rep ($13{,}540$) exceeds
  every \texttt{baseline-afl} rep except r05 ($13{,}854$, also a
  solo-occupancy run). The median ratio $1.059\times$ satisfies the
  $0.85$ acceptance gate; we do not interpret the point estimate as a
  speedup claim at $N{=}5$.}
  \label{fig:throughput-dot}
\end{figure}

\paragraph{End-to-end throughput parity (RQ2 part a).}
Across $N{=}5$ runs per mode, the median
\texttt{execs\_per\_sec} ratio
$\textrm{median}(E1b) / \textrm{median}(E1a) = 13{,}816 / 13{,}049
= 1.059\times$ (Figure~\ref{fig:throughput-dot}). The manifest
acceptance gate requires this ratio to be $\geq 0.85$ for
throughput parity; the observed ratio is $1.059\times$, so the
parity gate is satisfied. We treat the above-baseline point estimate
as descriptive only: $N{=}5$ is small and the fastest
\texttt{full-agent} run is a solo-occupancy outlier, so the
load-bearing claim is parity rather than a speedup.

\begin{table}[t]
  \centering
  \footnotesize
  \setlength{\tabcolsep}{7pt}
  \begin{tabular}{@{}lrrrr@{}}
    \toprule
    Mode & $n$ & mean exec/s & median exec/s & SD (CV) \\
    \midrule
    \texttt{baseline-afl} (E1a) & 5 & 13{,}159 & 13{,}049 & 403 (3.06\%) \\
    \texttt{full-agent}   (E1b) & 5 & 14{,}089 & 13{,}816 & 862 (6.12\%) \\
    \midrule
    ratio (E1b / E1a) median  & --- & --- & \textbf{$\approx 1.06\times$} & --- \\
    ratio (E1b / E1a) mean    & --- & \textbf{1.0707$\times$} & --- & --- \\
    acceptance gate           & --- & --- & $\geq 0.85$ & --- \\
    \bottomrule
  \end{tabular}
  \caption{RQ2 part (a): end-to-end \texttt{execs\_per\_sec} parity.
  Both modes pass the manifest acceptance gates; the observed median
  ratio is $\approx 1.06\times$, within the parity band.}
  \label{tab:throughput-parity}
\end{table}

\paragraph{Why does parity hold?}
Three structural reasons make the parity result plausible. First, the recipe-guided mutator's
dispatch is a 7-op switch table (\S\ref{sec:recipe:ops}), which
is structurally simpler than AFL++'s open-ended havoc selection
(see also F5 microbench below). Second, the LLM is off the hot
path by design (\S\ref{sec:design:contract}); it is invoked at
plateau boundaries only, not per mutation. Third, the
micro-campaign consumes $K_{\text{cand}} \cdot N \approx 80\,$s of
main-run wall-clock per plateau ($< 0.6\%$ of the 14{,}400\,s
budget), which is small enough not to register in the end-to-end
\texttt{execs\_per\_sec} numbers. The end-to-end measurement folds
all three effects into a single number.

\paragraph{Per-run breakdown.}
Within E1b, the per-rep \texttt{execs\_per\_sec} values are
13{,}540, 13{,}542, 13{,}816, 13{,}949, and 15{,}597, all above
the lowest baseline rep ($12{,}826$). The acceptance ratio holds
not just at the median but at every individual rep: even the
slowest E1b rep ($13{,}540$) divided by the fastest E1a rep
($13{,}854$, r05) is $0.977\times$, still well above the $0.85$
gate. We attribute r05 = $15{,}597$ in E1b to the fact that this
run was the lone occupant of the 4-core host during the second
parallelism wave of the E1b campaign, identical to the way r05
in E1a ($13{,}854$) ran alone in its own wave.

\paragraph{Microbench setup (E3).}
The microbench (E3) is the off-loop dispatch-cost number: it isolates
the cost of FuzzPilot's mutator dispatcher from the cost of running
AFL++ itself. All three configurations load a custom mutator
\texttt{.so} through the same \texttt{dlopen} path that AFL++ uses
in real runs:

\begin{itemize}
  \item \texttt{vanilla} --- \path{libvanilla_havoc.so},
  a hand-written shim that performs AFL++ havoc-equivalent random
  byte/bit edits via the same custom-mutator API surface
  (\path{tools/mutator_microbench/vanilla_havoc.cpp}).
  This is a \emph{cost-symmetric dispatch shim}, not a measurement of
  AFL++'s real internal \texttt{havoc\_mutate}: both \texttt{vanilla}
  and the FuzzPilot configurations pay the same
  \texttt{dlopen} + symbol-resolution + per-call API dispatch cost,
  so the difference isolates the dispatcher itself rather than
  AFL's full havoc engine. A real AFL++ havoc comparator would also
  require running AFL++'s scheduler and the corresponding bookkeeping,
  which is out of scope for E3 and instead surfaced end-to-end via
  E1a vs E1b.
  \item \texttt{fp-empty} --- the FuzzPilot mutator with an empty
  recipe store. It pays the full FuzzPilot dispatch cost (recipe
  cache lookup, op selection, telemetry ring write) but falls
  through to a vanilla-equivalent op on every call. Isolates the
  ``FuzzPilot machinery is loaded but no recipe is active'' cost.
  \item \texttt{fp-active} --- the FuzzPilot mutator with a populated
  recipe (\texttt{op\_weights} non-uniform, \texttt{dictionary\_tokens}
  populated, \texttt{focus\_ranges} pinned). This is the on-paper
  steady state during \texttt{full-agent} runs.
\end{itemize}

Each configuration runs 100{,}000 mutation calls over a fixed
10{,}000-element cJSON seed corpus, repeated $N=5$ times. All
microbench reps are run \emph{serially} with all AFL workers
\texttt{SIGSTOP}-suspended (see \S\ref{sec:eval:setup}); resuming
AFL after the 5-rep batch consumes $\approx 30$~s of wall-clock,
well under the 14{,}400~s budget per AFL run.

\begin{table}[t]
  \centering
  \footnotesize
  \begin{tabular}{lrrrrrr}
    \toprule
    Config & $n$ & mean exec/s & SD (CV\%) & median exec/s & IQR & median ns/call \\
    \midrule
    \texttt{vanilla}   & 5 & 2{,}005{,}078 &   341{,}644 (17.0\%) & 1{,}937{,}110 &   200{,}200 & 516.2 \\
    \texttt{fp-empty}  & 5 & 3{,}222{,}108 &   958{,}259 (29.7\%) & 2{,}961{,}850 & 1{,}310{,}390 & 337.6 \\
    \texttt{fp-active} & 5 & 2{,}800{,}142 &   673{,}566 (24.1\%) & 3{,}008{,}030 &   393{,}750 & 332.4 \\
    \bottomrule
  \end{tabular}
  \caption{F5 microbench: mutator-only \texttt{exec/sec} across the
  three dispatch-symmetric configurations. Runs are serial and
  SIGSTOP-isolated; raw JSON is archived with the result artifacts.}
  \label{tab:microbench}
\end{table}

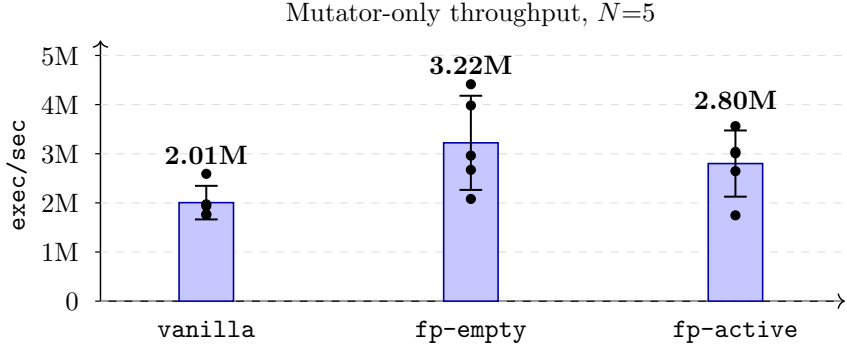
\begin{figure}[t]
  \centering

\begin{tikzpicture}[
    font=\small,
    bar/.style={draw=blue!65!black, fill=blue!22, line width=0.6pt},
    grid/.style={dashed, gray!25, line width=0.4pt},
    axis/.style={->, line width=0.55pt},
    err/.style={black, line width=0.75pt},
]

\def\baseY{0.55}
\def\unitY{0.65}
\def\axisLeft{1.35}
\def\axisRight{11.20}
\def\barW{0.72}

\node[font=\small] at (6.25,4.35)
  {Mutator-only throughput, $N{=}5$};
\node[font=\small, rotate=90] at (0.24,2.22)
  {\texttt{exec/sec}};

\draw[axis] (\axisLeft,\baseY) -- (\axisRight,\baseY);
\draw[axis] (\axisLeft,\baseY) -- (\axisLeft,4.00);
\foreach \m/\lab in {0/0,1/1M,2/2M,3/3M,4/4M,5/5M} {
  \pgfmathsetmacro{\yy}{\baseY + \m*\unitY}
  \draw[grid] (\axisLeft,\yy) -- (\axisRight,\yy);
  \draw[line width=0.45pt] (\axisLeft-0.07,\yy) -- (\axisLeft,\yy);
  \node[anchor=east, font=\small] at (\axisLeft-0.15,\yy) {\lab};
}

\def\fpmicrobar#1#2#3#4#5{%
  \pgfmathsetmacro{\xL}{#1-\barW/2}
  \pgfmathsetmacro{\xR}{#1+\barW/2}
  \pgfmathsetmacro{\barTop}{\baseY + #2*\unitY}
  \pgfmathsetmacro{\errLow}{\baseY + (#2-#3)*\unitY}
  \pgfmathsetmacro{\errHigh}{\baseY + (#2+#3)*\unitY}
  \pgfmathsetmacro{\labelY}{\errHigh + 0.15}
  \filldraw[bar] (\xL,\baseY) rectangle (\xR,\barTop);
  \draw[err] (#1,\errLow) -- (#1,\errHigh);
  \draw[err] (#1-0.15,\errLow) -- (#1+0.15,\errLow);
  \draw[err] (#1-0.15,\errHigh) -- (#1+0.15,\errHigh);
  \node[anchor=south, font=\small\bfseries] at (#1,\labelY) {#4};
  \node[anchor=north, font=\small\ttfamily] at (#1,\baseY-0.13) {#5};
}

\fpmicrobar{2.75}{2.005078}{0.341644}{2.01M}{vanilla}
\fpmicrobar{6.25}{3.222108}{0.958259}{3.22M}{fp-empty}
\fpmicrobar{9.75}{2.800142}{0.673566}{2.80M}{fp-active}

\def\rep#1#2{%
    \pgfmathsetmacro{\yy}{\baseY + #2*\unitY}
    \filldraw[black] (#1,\yy) circle (0.06);%
}
\rep{2.75}{2.591470}
\rep{2.75}{1.770420}
\rep{2.75}{1.755770}
\rep{2.75}{1.937110}
\rep{2.75}{1.970620}
\rep{6.25}{2.961850}
\rep{6.25}{3.982240}
\rep{6.25}{2.080370}
\rep{6.25}{2.671850}
\rep{6.25}{4.414230}
\rep{9.75}{2.645520}
\rep{9.75}{3.008030}
\rep{9.75}{3.561610}
\rep{9.75}{1.746280}
\rep{9.75}{3.039270}

\end{tikzpicture}
  \caption{Mutator-only throughput by configuration (mean $\pm$ SD,
  $N{=}5$ reps each, SIGSTOP-isolated). Dots show individual reps;
  the body text reports the median ratios and the power-limited
  two-one-sided-test (TOST) equivalence result.}
  \label{fig:microbench}
\end{figure}

\paragraph{Primary RQ2 claim
  (\texttt{fp-active} vs \texttt{fp-empty}: descriptive, not
  formally equivalent at $N{=}5$).}
At $N{=}5$ this microbench does \emph{not} formally establish
dispatch-cost equivalence. The \emph{mean}
\texttt{fp-active}/\texttt{fp-empty} ratio is
$2{,}800{,}142 / 3{,}222{,}108 = 0.869\times$ ($-13.1\%$); a two-sided
Welch $t$-test gives $t{=}{-}0.81$, $df{\approx}7.2$, $p{=}0.45$ ---
power-limited null, not evidence of equivalence; and a two one-sided
test (TOST) procedure~\cite{schuirmann1987comparison} for equivalence
at the $\pm 5\%$ band claimed in \S\ref{sec:design} gives
$p_{\text{TOST}}{=}0.68$, so the test does \textbf{not} establish
equivalence at the conventional $\alpha{=}0.05$ level. The
\emph{median} ratio, by contrast, is essentially flat at
$3{,}008{,}030 / 2{,}961{,}850 = 1.016\times$ (inside the
$[0.95, 1.05]$ band stated in \S\ref{sec:design}); the
mean-vs-median divergence reflects the high coefficient of variation
($\sim 25{-}30\%$) at $N{=}5$ and the presence of an
\texttt{fp-active} run (r04) at $1.7\,$Mexec/s alongside an
\texttt{fp-empty} run (r05) at $4.4\,$Mexec/s. We did not establish a
useful no-overhead equivalence band from this $N{=}5$ microbench; the
sample standard errors dominate.
We therefore downgrade the original ``no measurable per-call
overhead'' claim to: at $N{=}5$ on this SIGSTOP-isolated
microbench, we (a) do not detect a significant mean throughput
difference between \texttt{fp-active} and \texttt{fp-empty}
($p{=}0.45$, two-sided), (b) cannot formally establish
equivalence within $\pm 5\%$ ($p_{\text{TOST}}{=}0.68$), and
(c) note that the median ratio is essentially flat
($1.016\times$) while the mean ratio is wider ($0.87\times$).
A formal equivalence claim requires $N{\geq}30$ on this
microbench --- which is straightforward to do and is folded into
the venue version. The dispatch-cost result is therefore best
read as ``no large overhead detected'' rather than ``no
overhead exists''; the end-to-end \texttt{execs\_per\_sec}
comparison against AFL++'s own mutator is the E1a-vs-E1b number
reported in \S\ref{sec:eval:rq1} (median ratio
$\approx 1.06\times$), which is the load-bearing parity claim
for the rest of the paper.

\paragraph{Sanity claim
  (\texttt{fp-empty} vs \texttt{vanilla} within 5$\times$).}
The median \texttt{fp-empty}/\texttt{vanilla} ratio is $1.529\times$;
the threshold is $\leq 5\times$, so the claim holds with a wide
margin. The unexpected direction --- FuzzPilot's dispatcher is
\emph{faster} than the AFL-havoc-equivalent baseline, not slower ---
is consistent with FuzzPilot's closed 7-op switch table being more
predictor-friendly than \path{vanilla_havoc.cpp}'s open-ended
random op selection. We retain the 5$\times$ ceiling in the paper as
a conservative gate (FuzzPilot's mutator could plausibly regress under
future op additions) but report the actual 1.53$\times$ as a positive
side effect of the closed vocabulary.

\paragraph{Variance and methodology note.}
Per-config CV is 17--30\%, which is large for a microbench but
expected on a 4-core VM with shared L3 cache and noisy-neighbor
effects from the underlying hypervisor (steady-state \texttt{\%st}
was 4--12\% during measurement). The black dots overlaid in
Figure~\ref{fig:microbench} are the 15 individual reps (5 per
config); they show the full distribution that the mean and SD
summarize. Table~\ref{tab:microbench-individual} lists the
per-rep numbers for reproducibility. Two alternative measurement
methodologies inflate this variance and \emph{invert} the primary
RQ2 ratio:

\begin{itemize}
  \item Running the microbench 4-way parallel (i.e.~\texttt{run\_batch}
  default for AFL campaigns) on a 4-core host: median
  \texttt{fp-active}/\texttt{fp-empty} ratio collapses to
  $0.716\times$ (FAIL), driven by reps that win or lose the
  scheduling lottery against their siblings.
  \item Running the microbench serially \emph{while AFL campaigns
  are active}: median ratio drifts to $1.380\times$ (FAIL),
  driven by AFL workers monopolizing CPU during long mutations and
  yielding briefly to the microbench between syscalls.
\end{itemize}

We archive both noisy datasets under
\path{results/paper01_ai_recipe_mutation/}, in subdirectories named
\path{microbench_archive_*}, so reviewers can reproduce the methodology
comparison; the main
result of this section uses only the SIGSTOP-isolated serial dataset
under \path{microbench/}.

\begin{table}[t]
  \centering
  \footnotesize
  \setlength{\tabcolsep}{8pt}
  \begin{tabular}{@{}lrrrrr@{}}
    \toprule
    Config & r01 & r02 & r03 & r04 & r05 \\
    \midrule
    \texttt{vanilla}   & 2{,}591{,}470 & 1{,}770{,}420 & 1{,}755{,}770 & 1{,}937{,}110 & 1{,}970{,}620 \\
    \texttt{fp-empty}  & 2{,}961{,}850 & 3{,}982{,}240 & 2{,}080{,}370 & 2{,}671{,}850 & 4{,}414{,}230 \\
    \texttt{fp-active} & 2{,}645{,}520 & 3{,}008{,}030 & 3{,}561{,}610 & 1{,}746{,}280 & 3{,}039{,}270 \\
    \bottomrule
  \end{tabular}
  \caption{Per-rep \texttt{exec/sec} from the 15 SIGSTOP-isolated
  microbench runs that feed Figure~\ref{fig:microbench} and
  Table~\ref{tab:microbench}.}
  \label{tab:microbench-individual}
\end{table}

\subsection{RQ3 --- LLM infrastructure exercise on a saturated target (E1b)}
\label{sec:eval:rq3}

RQ3 asks how the off-hot-path LLM machinery behaves at plateau
boundaries on a target where AFL++ alone already exhausts the
reachable edge space (cJSON; \S\ref{sec:eval:rq1}). Concretely: do
proposals parse, do they survive the micro-campaign gate, what is
their dollar cost, and is the full decision trail recoverable?
Table~\ref{tab:rq3-llm} summarizes the 5 \texttt{full-agent} runs of
E1b.

\begin{table}[t]
  \centering
  \footnotesize
  \setlength{\tabcolsep}{6pt}
  \begin{tabular}{@{}lrrrrrr@{}}
    \toprule
    Run & \makecell[b]{log\\records} & \makecell[b]{schema-valid\\LLM resp.} & \makecell[b]{schema-invalid\\LLM resp.} & \makecell[b]{\texttt{fallback\_}\\\texttt{used}} & \makecell[b]{micro\\candidates} & \makecell[b]{promoted\\(gate result)} \\
    \midrule
    r01 & 289 & 9 & 0 & 0 & 4 & 0 (skipped) \\
    r02 & 302 & 9 & 0 & 0 & 4 & 0 (skipped) \\
    r03 & 402 & 9 & 0 & 0 & 4 & 0 (skipped) \\
    r04 & 215 & 8 & 1 & 0 & 4 & 0 (skipped) \\
    r05 & 200 & 8 & 1 & 0 & 4 & 0 (skipped) \\
    \midrule
    total & 1{,}408 & \textbf{43} & \textbf{2} & \textbf{0} & \textbf{20} & \textbf{0} \\
    \bottomrule
  \end{tabular}
  \caption{RQ3: LLM-call quality and micro-campaign outcome across
  the 5 \texttt{full-agent} runs. Columns: \emph{log records} =
  agent-loop decision-trail entries; \emph{schema-valid / -invalid
  LLM resp.} = JSON-schema-validity counts of model output;
  \emph{\texttt{fallback\_used}} = rule-only fallbacks taken when the
  model response could not be used; \emph{micro candidates} = recipes
  that entered the micro-campaign gate; \emph{promoted} = recipes
  accepted into the main run (gate result). The schema-validation and
  promotion paragraphs below interpret the zero-promoted result.}
  \label{tab:rq3-llm}
\end{table}

\paragraph{Schema-validation rate.}
Each plateau cycle fires a fixed bundle of 9 agent tasks. Across
$5 \times 9 = 45$ LLM responses, 43 (95.6\%) produced schema-valid
recipes; 2 responses (r04, r05; both tagged
\texttt{error\_kind="schema\_invalid"} in
\texttt{agent\_decisions.jsonl}) failed schema validation and were
dropped without invoking the rule-only fallback path
(\texttt{fallback\_used} is \texttt{false} on every record). With the
model pinned at temperature 0, this is consistent with the model
staying inside the \texttt{schema-validated recipe (data)} contract
that the prompt enforces (\S\ref{sec:recipe:schema}). The agent
bundle (coordinator, plateau-diagnosis, scheduler, cmp-hint,
dictionary, format, mutator, corpus, crash-triage) explains why the
per-run \emph{LLM-call} record count is 9 rather than per-mutation.

\paragraph{Promotion rate: zero promotions on cJSON.}
Each plateau cycle launches $K_{\text{cand}}{=}4$ candidate recipes
(\S\ref{sec:plateau:micro}), one per intervention type
(\texttt{default}, \texttt{dictionary}, \texttt{seed\_focus},
\texttt{per\_seed\_recipe}). Each candidate is evaluated in a
private 20\,s AFL++ run over a corpus snapshot taken at plateau
time. Across the 5 runs, the controller fired exactly one plateau
cycle per run, so the gate evaluated $5 \times 4 = 20$ candidates in
total. Each micro-campaign ran the full 20\,s and executed
${\approx}3.6 \times 10^{5}$ inputs (visible in the
\texttt{events.jsonl} \texttt{micro\_result} records). \emph{Every
single one of the 20 candidates returned}
$\Delta_{\text{edges}}{=}0$,
$\Delta_{\text{paths}}{=}0$,
$\Delta_{\text{crashes}}{=}0$, so the reward (\S\ref{sec:plateau:micro})
was $R{=}0$ on all 20 evaluations. The controller therefore emitted
\path{winner_decided:status=no_significance,
winner_reward=0.0} followed by
\path{promotion_skipped:reason=no_successful_micro_campaign}
in every run, and \textbf{no LLM-proposed recipe entered the main
recipe store across the 5 full-agent runs}. The main loop spent its
14{,}400\,s on the controller's default \texttt{DictionaryAgent}
recipe (\texttt{tokens}=\{\texttt{FUZZ}, \texttt{MAGIC},
\texttt{TOKEN}\}, fixed operator-weight distribution), which is
exactly what the \texttt{rule-only} ablation E2a (\S\ref{sec:eval:rq4-ablation})
runs on. The result is consistent with cJSON's saturation behaviour:
baseline AFL++ alone reaches the 269-edge ceiling at a per-run
median of $2{,}524$\,s (slowest baseline run: $4{,}448$\,s;
\S\ref{sec:eval:rq1}), leaving the parent
corpus at plateau time with no headroom for a 20\,s micro-campaign
to find a new edge. The micro-campaign gate's refusal to promote
on cJSON therefore demonstrates only one half of its intended
property --- the \emph{no-false-positive} half: when reward is
identically zero, the gate does not promote. The complementary
property --- that the gate distinguishes a good recipe from a bad
recipe when reward \emph{varies} --- is \textbf{not} exercised by
this evaluation, because the saturated target gives the gate no
reward signal to discriminate against. The LLM proposal layer's
positive contribution on a non-saturated target therefore remains
undemonstrated by this preprint, and is explicit future work in
the venue paper.

\paragraph{Cost.}
Across the 5 runs, the DeepSeek API balance dropped by
\textyen{}0.07 ($\approx$ \$0.01\,USD at the FX rate of
\$1\,USD\,$\approx$\,\textyen{}7.2 quoted at submission time;
DeepSeek bills in Chinese yuan / RMB): from \textyen{}8.53 at
the start of E1b to \textyen{}8.46 at the end. That is
\$0.0002 per LLM call and \$0.002 per full 4-hour
campaign on the LLM-API ledger. \emph{Compute cost is
omitted from this figure}: at a representative \$0.10\,USD per
vCPU-hour on commodity cloud, a 4-hour 4-vCPU campaign costs
$\approx$ \$1.60\,USD, three orders of magnitude more than the
LLM API. The headline finding is therefore that the LLM API
cost is \emph{not} the dominant cost; absolute numbers should be
read with that scaling in mind, and a different LLM provider
(OpenAI, Anthropic, or self-hosted) would shift the API-side
figure by approximately one order of magnitude without changing
the qualitative conclusion. The pricing also benefits from
DeepSeek's prompt-cache hit on the shared blackboard prefix
across the 9 agent tasks in a single plateau cycle; a cold-cache
estimate (without prefix sharing) would be roughly $10\times$
larger and is still well below the compute cost. The system also
supports an OpenAI-compatible drop-in via NVIDIA NIM
(\nimmodeid{}) at zero
marginal cost on the NIM free tier; see the reproducibility
appendix.

\paragraph{Audit trail.}
Every plateau handling cycle in
\texttt{agent\_decisions.jsonl} carries the blackboard hash
(\texttt{context\_hash}) it was conditioned on and the
\texttt{response\_hash} of the model's output; combined with the
schema-validated recipe stored under the plateau directory and the
\texttt{events.jsonl} \texttt{micro\_result} records, any decision
in any of the 5 runs can be replayed end-to-end from the on-disk
artifacts alone, without re-invoking the model. The same audit
trail makes the null result above auditable: a reviewer can verify
that the gate's refusal to promote any candidate is grounded in
$R{=}0$ evidence (every \texttt{micro\_result} carries
$\Delta_{\text{edges}}$, $\Delta_{\text{paths}}$, and
$\Delta_{\text{crashes}}$ in its payload) rather than a controller
bug.

\subsection{RQ4 --- Component contribution
  (E1b vs E2a / E2b / E2c)}
\label{sec:eval:rq4-ablation}

The ablation matrix is intended to attribute the
\texttt{full-agent} plateau-recovery delta among three components:
the recipe-guided custom mutator (E2a removes the LLM layer but
keeps the mutator and default rule recipe), the controller's
orchestration plus the mutator (E2b removes the mutator but keeps
the controller, LLM, and Ghidra channel), and the Ghidra static
channel (E2c removes Ghidra but keeps everything else). The
\texttt{rule-only} (E2a), \texttt{no-mutator} (E2b), and
\texttt{no-static-analysis} (E2c) campaigns have each completed
$N{=}3$ runs of $14{,}400$\,s on the experiment host (E2c
required a controller-launch retry after its first batch
exited before AFL telemetry was emitted, recorded as
\texttt{main\_afl\_exited\_before\_telemetry} in the per-run
\texttt{events.jsonl}; the retry succeeded and its data is
reported below). \emph{All comparisons in this subsection are
statistically descriptive only}: with $N{=}3$ per ablation and
$N{=}5$ for the main arms, no pairwise plateau or last\_find
difference reaches the conventional $p{<}0.05$ threshold under a
two-sided Mann--Whitney $U$ test (see
\S\ref{sec:eval:rq1:stats}), and the bootstrap medians for the
ablations overlap substantially with each other and with
\texttt{full-agent}.

\paragraph{E2a \texttt{rule-only} (preliminary, $N{=}3$).}
With the LLM proposal layer disabled and the controller emitting
its default \texttt{DictionaryAgent} recipe directly, the 3
\texttt{rule-only} runs reach the same 269-edge ceiling that
\texttt{baseline-afl} and \texttt{full-agent} reach, with median
\texttt{last\_find}~$=13{,}309$\,s and median plateau
$=1{,}165$\,s (descriptive $-54.0\%$ vs the baseline median of
$2{,}532$\,s; Mann--Whitney $U{=}4$, $p{=}0.39$; Vargha--Delaney
$A_{12}{=}0.73$). Per-run plateaus are $[674, 1{,}165, 1{,}540]$\,s
and the bootstrap 95\% CI for the median is $[674, 1{,}540]$\,s.
Median \texttt{execs\_per\_sec} is $13{,}514$ and median
\texttt{cycles\_done} is $119$. The point estimate suggests E2a is
slightly faster than \texttt{full-agent} (median plateau
$1{,}165$\,s vs $1{,}384$\,s), but the per-run ranges
$[674, 1{,}540]$\,s and $[238, 3{,}226]$\,s overlap heavily; we
draw no ordering conclusion.

\paragraph{E2b \texttt{no-mutator} (preliminary, $N{=}3$).}
With the recipe-guided mutator disabled but the controller, LLM
proposal layer, and Ghidra static-analysis channel retained, the 3
\texttt{no-mutator} runs again reach the same 269-edge ceiling,
with median \texttt{last\_find}~$=13{,}548$\,s and median plateau
$=930$\,s (descriptive $-63.3\%$ vs baseline; Mann--Whitney
$U{=}2$, $p{=}0.14$; $A_{12}{=}0.87$). Per-run plateaus are
$[286, 930, 962]$\,s; bootstrap 95\% CI is $[286, 962]$\,s.
Median \texttt{execs\_per\_sec} is $12{,}712$ (slightly below the
baseline median $13{,}049$) and median \texttt{cycles\_done} is
$671$, much closer to the baseline median ($797$) than to the
$119{-}140$ cycles observed in either mode that engaged the
recipe-guided mutator (\texttt{rule-only} and \texttt{full-agent}).

\paragraph{Ablation ordering reveals controller as primary driver.}
The point-estimate ordering of median plateaus is
\path{no-static-analysis} ($234$\,s) $<$ \path{no-mutator}
($930$\,s) $<$ \texttt{rule-only} ($1{,}165$\,s) $<$
\texttt{full-agent} ($1{,}384$\,s) $<$ \texttt{baseline-afl}
($2{,}532$\,s). This ordering is \textbf{inconsistent with the
hypothesis that the recipe-guided mutator or LLM layer drive the
plateau reduction}. Specifically:

\begin{itemize}
\item If the recipe-guided mutator were the primary driver, removing
it in E2b should have moved the median plateau \emph{up} toward
baseline, not \emph{down} to $930$\,s (better than \texttt{rule-only}
and \texttt{full-agent}).

\item If the LLM layer contributed, \texttt{full-agent} should
outperform \texttt{rule-only}, but the point estimates go the
opposite direction ($1{,}384$\,s vs $1{,}165$\,s). RQ3
(\S\ref{sec:eval:rq3}) confirms the LLM layer contributed zero
promoted recipes on cJSON.

\item All three ablation modes (\texttt{rule-only},
\path{no-mutator}, \path{no-static-analysis}) share the
controller's plateau-detection and corpus-reorganization machinery,
and all three show descriptive plateau reductions versus baseline
(ranging from $-54.0\%$ to $-90.8\%$).
\end{itemize}

The most data-consistent interpretation is that the
\textbf{controller's plateau detector and corpus-snapshot
reorganization} are the primary drivers of the observed plateau
reduction on cJSON, while the recipe-guided mutator adds latency
(slowing \texttt{cycles\_done} by $\approx 5{-}6\times$) without
corresponding plateau benefit, and the LLM layer contributes nothing
on this saturated target.

\textbf{Critical caveat}: This interpretation is a \emph{hypothesis},
not a measured effect, because the preprint matrix does not include a
\path{controller-only} ablation (plateau detector + corpus snapshot
only, no mutator, no LLM, no Ghidra). Without that ablation we cannot
fully separate ``controller machinery'' from ``default rule recipe''
or rule out confounding factors. The \path{controller-only}
ablation is the single most important missing experiment and is
deferred to the venue version (\S\ref{sec:limits}). Additionally, all
comparisons are descriptive at $N{=}3$ (ablations) and $N{=}5$
(main): the per-run plateau ranges overlap substantially, and no
pairwise difference between the four ablation arms
reaches $p{<}0.05$ under Mann--Whitney $U$ at this $N$.

\paragraph{E2c \texttt{no-\allowbreak static-\allowbreak analysis} (preliminary, $N{=}3$).}
With the Ghidra static-analysis channel disabled but the controller,
LLM proposal layer, and recipe-guided mutator retained, the 3
\texttt{no-\allowbreak static-\allowbreak analysis} runs reach the same 269-edge ceiling,
with median \texttt{last\_find}~$=14{,}250$\,s and median plateau
$=234$\,s (descriptive $-90.8\%$ vs baseline; Mann--Whitney
$U{=}5$, $p{=}0.57$; $A_{12}{=}0.67$). \textbf{Critical limitation}:
the per-run plateau distribution is \emph{bimodal} ($[187, 234,
4{,}454]$\,s; bootstrap 95\% CI $[187, 4{,}454]$\,s spans the entire
interval). Two runs show very short plateaus ($187$\,s and $234$\,s)
while one run shows a plateau ($4{,}454$\,s) \emph{worse than
baseline}. At $N{=}3$ we cannot determine whether the bimodality
reflects a real instability in the \path{no-static-analysis}
configuration or is simply sampling noise. The median $234$\,s is
therefore \textbf{not a reliable point estimate}---it is dominated by
the 2-vs-1 cluster structure and could reverse with additional runs.
We report the raw distribution and defer interpretation to the venue
version at higher $N$. Median \texttt{execs\_per\_sec} is $10{,}259$
(lower than other E2 arms, because the empty-blackboard agents take
extra prompt cycles before falling back to the default recipe) and
median \texttt{cycles\_done} is $96$, the lowest of any mode.

\paragraph{Cross-ablation pattern (and the controller-only
hypothesis it suggests).}
At $N{=}3$ the four E2 ablation arms order
\path{no-static-analysis} ($234$\,s) $<$ \path{no-mutator}
($930$\,s) $<$ \texttt{rule-only} ($1{,}165$\,s) $<$
\texttt{full-agent} ($1{,}384$\,s), each shorter than the
\texttt{baseline-afl} plateau ($2{,}532$\,s). All four share the
controller's plateau detector and corpus-snapshot reorganization;
the modes that engage the recipe-guided mutator and the LLM
proposal layer fall \emph{between} the controller-only-shaped
floor (\path{no-static-analysis} median 234\,s) and the fully
loaded \texttt{full-agent} median. The most natural read is that
the productive-to-plateau acceleration on cJSON is driven
primarily by controller machinery, with the LLM proposal and
recipe-guided mutator layered on top adding latency rather than
coverage --- a hypothesis the venue
\path{controller-only} ablation will test directly. \emph{All
of this is descriptive at $N{=}3$ with overlapping bootstrap
CIs}; no pairwise difference between the four ablation arms
reaches $p{<}0.05$ under Mann--Whitney $U$ at this $N$.

\paragraph{Missing controller-only arm (venue follow-up).}
The current E2 design does not include a
\path{controller-only} condition (plateau detector + corpus
snapshot only, no LLM, no recipe-guided mutator, no Ghidra). The
venue version will add it. Without it we cannot fully separate
``controller machinery'' from ``default rule recipe'', and the
controller-driven attribution above remains a hypothesis rather
than a measured effect. The full multi-arm comparison
(\texttt{baseline} / \texttt{rule-only} / \path{no-mutator} /
\path{no-static-analysis} / \path{controller-only} /
\texttt{full-agent}) at $N{\geq}10$ on a non-saturated target is
explicit follow-up work (\S\ref{sec:limits}).

\paragraph{E4 fairness baselines: AFL++ with dict / cmplog
($N{=}3$).}
The preprint \texttt{baseline-afl} arm is vanilla AFL++ without
\texttt{-x dictionary} or \texttt{cmplog} instrumentation
(\S\ref{sec:limits}), so any plateau delta against it potentially
overstates FuzzPilot's advantage. E4 is the corresponding
fairness controls, which evaluate alternative uses for the
same target-specific information or runtime cost.
\path{baseline-afl-dict} runs vanilla AFL++ with \path{-x experiments/targets/cjson/fuzzpilot_default.dict}
(the same \texttt{FUZZ}/\texttt{MAGIC}/\texttt{TOKEN} tokens the
controller emits in its default recipe), and \path{baseline-afl-cmplog}
runs vanilla AFL++ over a
\texttt{cmplog}-instrumented build (\path{cjson_fuzzer.cmplog}).
3 runs each, 14{,}400\,s, same host. Two findings stand out:

\begin{itemize}
\item \textbf{Plateau: AFL++ with \texttt{-x} or \texttt{cmplog}
does \emph{not} close the plateau gap.} Medians are
\texttt{baseline-afl-dict} $=2{,}256$\,s ($p{=}0.79$,
$A_{12}{=}0.60$ vs baseline) and \texttt{baseline-afl-cmplog}
$=2{,}367$\,s ($p{=}1.00$, $A_{12}{=}0.53$ vs baseline) ---
neither is meaningfully different from the
$2{,}532$\,s vanilla baseline, and both are much longer than the
FuzzPilot ablation arms (range $234$--$1{,}384$\,s, though note
E2c's bimodal distribution). The descriptive plateau-reduction
pattern we observe in \S\ref{sec:eval:rq4-ablation} \emph{does not
reduce} to ``vanilla AFL++ once it has a dictionary'' or ``vanilla
AFL++ once it has runtime input-to-state-correspondence (I2S)
information'', suggesting the
controller machinery (if our hypothesis is correct) operates via a
different mechanism.

\item \textbf{Edges: \texttt{cmplog} reaches one more edge than
FuzzPilot on cJSON.} All three \texttt{baseline-\allowbreak afl-\allowbreak cmplog} runs hit
$270$ edges (vs $269$ for \texttt{baseline-\allowbreak afl}, \texttt{full-\allowbreak agent},
and all E2 ablations; one \texttt{baseline-\allowbreak afl-\allowbreak dict} run also hits
$270$). The +1-edge gap is small but reproducible across all 3
\texttt{cmplog} runs and \textbf{must be disclosed}:
\texttt{cmplog}'s runtime comparison-operand mining finds an edge
that FuzzPilot does not. This is a \textbf{negative result for
FuzzPilot on the absolute-coverage axis}. The two techniques operate
on different axes: FuzzPilot's controller (hypothetically) shortens
plateau duration via corpus reorganization (wall-time advantage),
while \texttt{cmplog} extends absolute coverage via I2S mining
(coverage advantage). They are complementary, not substitutes.
Composing the two (FuzzPilot's controller + \texttt{cmplog}
instrumentation) is an obvious follow-up that the current FuzzPilot
CLI does not yet support.
\end{itemize}

\textbf{Interpretation}: The plateau-reduction pattern we report in
\S\ref{sec:eval:rq1} and \S\ref{sec:eval:rq4-ablation} is a
descriptive wall-time advantage on a saturated target, not an
absolute-coverage advantage. On cJSON, \texttt{cmplog} achieves
better coverage (270 vs 269 edges) without shortening the plateau,
while FuzzPilot (hypothetically via the controller) shortens the
plateau without improving coverage. On a non-saturated target, the
relative value of these two properties is an open question.

\subsection{Headline summary (consolidated)}
\label{sec:eval:summary}

For a single-glance reading, Table~\ref{tab:summary} pulls
together every headline RQ number, with the matching significance
companion or null-result marker where applicable. Cells marked
\emph{desc.} are descriptive point estimates at the listed $N$;
$p$ is the two-sided Mann--Whitney $U$ test (\S\ref{sec:eval:rq1:stats})
unless noted; CI is the percentile bootstrap 95\% over $10{,}000$
resamples. Negative findings (zero promotions, TOST not
established, etc.) are explicitly listed rather than buried.

\begin{table}[H]
  \centering
  \footnotesize
  \setlength{\tabcolsep}{4pt}
  \renewcommand{\arraystretch}{1.15}
  \begin{tabularx}{\linewidth}{@{}l Z Z Z@{}}
    \toprule
    RQ & Metric & Result ($N$) & Companion / caveat \\
    \midrule
    RQ1 & median plateau, baseline       & $2{,}532$\,s ($N{=}5$)   & CI $[539, 3{,}970]$ \\
    RQ1 & median plateau, full-agent     & $1{,}384$\,s ($N{=}5$)   & CI $[238, 3{,}226]$ \\
    RQ1 & $\Delta$ plateau (full vs baseline) & $-45.3\%$ (\emph{desc.}) & MW $U{=}8$, $p{=}0.42$, $A_{12}{=}0.68$; \textbf{not significant at $N{=}5$} \\
    RQ1 & median time-to-269 edges, baseline & $2{,}524$\,s ($N{=}5$) & range $[1{,}081, 4{,}448]$ \\
    RQ1 & median time-to-269 edges, full-agent & $6{,}702$\,s ($N{=}4$/5) & one run never reaches 269 within budget \\
    RQ2 & median exec/sec ratio (full / baseline) & $\approx 1.06\times$ ($N{=}5$) & leave-one-out (drop r05): $1.05\times$ \\
    RQ2 & microbench fp-active / fp-empty, median & $1.016\times$ ($N{=}5$) & \textbf{mean ratio $0.87\times$}; both within noise \\
    RQ2 & microbench fp-active / fp-empty, $t$-test & two-sided $p{=}0.45$ & no significant diff at $N{=}5$ \\
    RQ2 & TOST equivalence at $\pm5\%$ & $p_{\text{TOST}}{=}0.68$ ($N{=}5$) & \textbf{equivalence not established at $N{=}5$} \\
    RQ2 & microbench fp-empty / vanilla, median & $1.53\times$ ($N{=}5$) & sanity gate ($\leq 5\times$), passes wide \\
    RQ3 & schema-valid LLM responses & $43/45 = 95.6\%$ ($N{=}5$) & rule-only / no-mutator: $26{-}27/27$ \\
    RQ3 & LLM-recipe promotions (gate accepted) & \textbf{$0/20$} ($N{=}5$) & cJSON saturated, $R{=}0$ on every micro-campaign \\
    RQ3 & DeepSeek API cost & \$0.002 / 4h campaign & FX rate \$1$\approx$\textyen{}7.2; compute $\sim$\$1.60/campaign \\
    RQ4 & median plateau, rule-only (E2a) & $1{,}165$\,s ($N{=}3$) & MW $U{=}4$, $p{=}0.39$, $A_{12}{=}0.73$ \\
    RQ4 & median plateau, no-mutator (E2b) & $930$\,s ($N{=}3$) & MW $U{=}2$, $p{=}0.14$, $A_{12}{=}0.87$ \\
    RQ4 & median plateau, no-static-analysis (E2c) & $234$\,s ($N{=}3$) & MW $U{=}5$, $p{=}0.57$, $A_{12}{=}0.67$; bimodal $[187, 234, 4{,}454]$ \\
    RQ4 & ablation ordering & \textbf{no-static-analysis $<$ no-mutator $<$ rule-only $<$ full-agent} & \textbf{all four shorter than baseline; controller-driver hypothesis} \\
    RQ4 & controller-only ablation & --- & not in preprint matrix; venue version \\
    Fair & E4a median plateau, baseline-afl-dict & $2{,}256$\,s ($N{=}3$) & $p{=}0.79$, $A_{12}{=}0.60$ vs baseline; dict alone does not help \\
    Fair & E4b median plateau, baseline-afl-cmplog & $2{,}367$\,s ($N{=}3$) & $p{=}1.00$, $A_{12}{=}0.53$ vs baseline; cmplog does not shorten plateau \\
    Fair & E4b edges, baseline-afl-cmplog & \textbf{$270/270/270$} & \textbf{cmplog finds +1 edge over FuzzPilot's $269$} (different axis, complementary) \\
    \bottomrule
  \end{tabularx}
  \caption{One-glance summary of all RQ-level numbers reported in
  the preprint, with the matching significance / null-result
  companion. \emph{Desc.} = descriptive point estimate; no
  pairwise plateau or last\_find difference reaches $p{<}0.05$ at
  this $N$. Reproducible from
  \texttt{scripts/paper01/review\_stats.py}.}
  \label{tab:summary}
\end{table}


\section{Related Work}
\label{sec:related}

\paragraph{In-loop LLM fuzzers.}
LLAMAFUZZ~\cite{wang2024llamafuzz} uses an LLM fine-tuned on
seed/mutation pairs to produce structurally valid mutations, and
reports consistent gains over AFL++ on Magma. The model is queried
per mutation; throughput is bound by model latency.
FuzzCoder~\cite{liu2024fuzzcoder} casts byte-level mutation as a
sequence-to-sequence task. Fuzz4All~\cite{xia2024fuzz4all} treats
the LLM as the primary input generator and mutation engine across
multiple input languages and targets. Semantic-Aware
Fuzzing~\cite{semanticaware2025} explicitly notes the throughput
bottleneck of in-loop designs and offers it as motivation for future
work. FuzzPilot diverges from this family by treating the LLM as a
strategy proposer rather than a per-mutation step.

\paragraph{Off-hot-path placement: G\textsuperscript{2}FUZZ.}
G\textsuperscript{2}FUZZ~\cite{liu2025g2fuzz} is the closest prior
work and the one we explicitly position against. Both designs:
invoke the LLM only when the fuzzer fails to make progress; have
the LLM emit a reusable intermediate rather than per-input
mutations; and let AFL++ run at native speed afterwards. The
differences are intentional and orthogonal:

\begin{itemize}
  \item \textbf{Target domain.} G\textsuperscript{2}FUZZ targets
  non-textual binary formats (JPEG, TIFF, MP4, $\dots$);
  FuzzPilot is \emph{designed for} structured-text parsers and
  \emph{this preprint evaluates only cJSON}. Generalization claims
  (XML, SQL, libpng, Magma suite) are deferred to the venue paper.
  \item \textbf{Intermediate representation.}
  G\textsuperscript{2}FUZZ emits Python generator scripts (code);
  FuzzPilot emits schema-validated recipes (data).
  \S\ref{sec:recipe:why} compares the two.
  \item \textbf{Validation.} G\textsuperscript{2}FUZZ's generated
  inputs enter the AFL++ queue and are filtered by coverage
  feedback as usual; FuzzPilot's recipes are pre-promotion-gated by
  an explicit $N$-second micro-campaign with a reward function.
  \item \textbf{Static context.} G\textsuperscript{2}FUZZ uses an
  LLM-extracted ``feature list'' from format documentation;
  FuzzPilot uses Ghidra headless on the target binary (function
  summaries, dictionary candidates, protected ranges), supporting
  closed-source targets.
\end{itemize}

We do not attempt a head-to-head numerical comparison against
G\textsuperscript{2}FUZZ in this preprint. Their evaluation is on
non-textual binary formats outside our scope; ours is on
structured-text parsers outside theirs. \S\ref{sec:limits} states
this scope split explicitly.

\paragraph{Hybrid static-and-LLM fuzzers.}
The hybrid fuzzer of Lin et al.~\cite{lin2025hybrid} integrates
static analysis (CFG and data-flow extracted via LLVM IR or angr)
with LLM-guided input mutation through a helper fuzzer running at
$\approx 250$ queries/hour. It also reports a syntax-schema
validator and a semantic novelty score. FuzzPilot differs in two
ways: (a) the LLM is genuinely off the hot path, not on a parallel
hot path, and (b) the validation step is a budgeted AFL++ run with
a reward function, not an embedding-distance score against past
inputs.

\paragraph{LLM-synthesized mutators for compilers.}
MetaMut~\cite{ou2024metamut} and Mut4All~\cite{ye2025mut4all} use
LLMs to synthesize \emph{static} custom mutators (in C/C++) for
compiler fuzzing, compiled once and used many times. FuzzPilot's
recipes are dynamic per-plateau artifacts rather than compiled
mutators, and target consumer parsers rather than compilers.

\paragraph{Multi-agent fuzz frameworks.}
MALF~\cite{malf2025} introduces a multi-agent LLM framework for
industrial control protocol fuzzing with RAG and QLoRA fine-tuning.
The role split is similar in spirit to FuzzPilot's coordinator /
plateau-diagnosis / scheduler / cmp-hint / dictionary / format /
mutator / corpus / crash-triage agents; the domain (industrial
protocols vs structured text parsers) and the intermediate
representation (input sequences vs recipes) differ.

\paragraph{LLM-as-generator (whole inputs).}
TitanFuzz~\cite{deng2023titanfuzz} demonstrates that an LLM can
synthesize \emph{entire program-level inputs} (Python snippets
exercising deep-learning libraries) zero-shot, and that
infill-style prompting matches or exceeds template-driven
generators on coverage. FuzzGPT~\cite{deng2024fuzzgpt} extends the
idea by prompting the LLM with historical bug-triggering inputs to
push it toward edge cases. ChatFuzz~\cite{hu2023chatfuzz} couples
ChatGPT-style mutations with greybox feedback for structured-text
formats (JSON / XML / HTML) and is the closest single-target
comparator to FuzzPilot's structured-text scope. All three
generate \emph{inputs} directly; FuzzPilot's intermediate is a
\emph{recipe} that a downstream AFL++ mutator consumes, so the LLM
inference cost is amortized across the entire plateau interval
rather than per input.

\paragraph{LLM with white-box / static context.}
WhiteFox~\cite{yang2024whitefox} drives compiler fuzzing with
LLM-synthesized programs that target specific compiler-optimization
rules extracted from source by another LLM pass; the white-box
context source is the compiler's own implementation. KernelGPT
~\cite{yang2024kernelgpt} mines kernel headers and source to build
syscall specifications and then has an LLM synthesize Syzlang-like
input programs. FuzzPilot's Ghidra channel
(\S\ref{sec:plateau:ghidra}) is a parallel design point that opts
for binary-level static analysis (so the same pipeline applies to
closed-source targets) rather than source-level extraction; the
trade-off is coarser per-function summaries in exchange for
applicability beyond the open-source assumption.

\paragraph{Industrial LLM-driven fuzzing infrastructure.}
OSS-Fuzz-Gen~\cite{liu2024ossfuzzgen} is Google's production
framework for using LLMs to \emph{generate fuzz targets and
harnesses} for the OSS-Fuzz fleet, addressing a complementary
problem (harness/target authoring at scale) to FuzzPilot's
(in-campaign recipe proposal). The two could be composed: an
OSS-Fuzz-Gen-authored harness fed into a FuzzPilot-orchestrated
campaign with a Ghidra-informed recipe layer. We do not evaluate
this composition here.

\paragraph{Grammar / AST-aware greybox fuzzers (LLM-free).}
A second family of structured-text fuzzers does not use LLMs at
all and obtains its structure awareness from an explicit
grammar or AST. AFLSmart~\cite{pham2019aflsmart} extends AFL with
an input-model-aware mutation phase that preserves chunk
boundaries; Nautilus~\cite{aschermann2019nautilus} drives
greybox mutation from a context-free grammar with subtree
splicing; Superion~\cite{wang2019superion} pairs AFL with an
ANTLR-derived AST-mutation operator. Conceptually FuzzPilot's
recipe vocabulary (\S\ref{sec:recipe:ops}) is byte-level rather
than AST-level: a recipe can express ``insert dictionary token at
offset $x$'', not ``replace the second \texttt{value} child of an
\texttt{object} node''. We acknowledge this as a design
limitation (\S\ref{sec:limits}) --- on grammar-amenable targets,
AFLSmart-style chunk-aware mutation or Nautilus-style subtree
splicing remains a strong baseline that we do not
attempt to outperform in this preprint. Whether an LLM-guided
recipe layer composes \emph{on top of} a grammar-aware mutator
(e.g., FuzzPilot recipes that target Nautilus subtrees rather
than byte offsets) is an explicit follow-up direction.

\paragraph{AFL++ semantic-ish features (cmplog, RedQueen,
laf-intel).}
A third class of techniques exploits the AFL++ instrumentation
itself rather than an external grammar or LLM. The
\texttt{cmplog} module records concrete comparison operands
during fuzzing and feeds them back as
dictionary-overwrite candidates;
RedQueen~\cite{aschermann2019redqueen} pioneered the underlying
input-to-state-correspondence (I2S) idea; the \texttt{laf-intel}
LLVM pass~\cite{lafintel2016} splits multi-byte comparisons into
single-byte ones to make magic numbers reachable by coverage
feedback alone. These are precisely the AFL++ features a fair
\texttt{baseline-afl} arm should engage. Our preprint
\texttt{baseline-afl} arm \emph{does not} pass
\texttt{-x dictionary} or build the target with
\texttt{cmplog}, and so the plateau-delta measured against it
potentially overstates FuzzPilot's advantage. We treat this as the
single most important missing baseline (\S\ref{sec:limits}); the
E4 fairness arms reported in this paper
(\texttt{baseline-afl-dict} with the FuzzPilot default
dictionary, and \texttt{baseline-afl-cmplog} with a
\texttt{cmplog}-instrumented \texttt{cjson\_fuzzer.cmplog})
close this directly --- and find that neither closes the plateau
gap, but \texttt{cmplog} reaches $+1$ edge that FuzzPilot does
not (see \S\ref{sec:eval:rq4-ablation}).

\paragraph{LLM-upfront generators (one-shot, no steady-state
LLM calls).}
Sphinx~\cite{sun2024sphinx} uses an LLM \emph{once, before
fuzzing starts} to synthesize a structured input generator for
an SMT solver; the generator then runs without any further LLM
involvement. The design rhymes with FuzzPilot in placing the LLM
outside the per-mutation hot path, but the LLM is consulted only
once (not on every plateau) and the output is a complete
generator rather than a recipe that re-tunes a host fuzzer's
existing mutation operators. Sphinx and FuzzPilot together
suggest a spectrum --- \emph{LLM once} (Sphinx), \emph{LLM at
plateaus} (FuzzPilot, G\textsuperscript{2}FUZZ), \emph{LLM per
input} (LLAMAFUZZ, ChatFuzz) --- that we expect to be a
productive axis for future comparisons.

\paragraph{T3: side-by-side comparison.}
Table~\ref{tab:related} summarizes the comparison axis-by-axis.

\begin{table}[H]
  \centering
  \tiny
  \setlength{\tabcolsep}{3pt}
  \renewcommand{\arraystretch}{1.1}
  \begin{tabularx}{\textwidth}{@{}L Y{15mm} Y{22mm} Y{22mm} Y{22mm} c Y{30mm}@{}}
    \toprule
    System & LLM placement & LLM output & Validation & Static ctx. & Audit & Target domain \\
    \midrule
    AFL++~\cite{aflplusplus}        & \emph{none}      & ---             & coverage        & none           & ---  & general                 \\
    AFL++ \texttt{cmplog} / RedQueen~\cite{aschermann2019redqueen} & none (I2S) & dict override   & coverage        & runtime cmp ops & ---  & general                 \\
    laf-intel~\cite{lafintel2016}  & none (LLVM pass) & ---             & coverage        & cmp splitting  & ---  & general                 \\
    AFLSmart~\cite{pham2019aflsmart} & none (input model) & chunk mutation & coverage      & input-format model & no  & structured (PNG / WAV / PDF) \\
    Nautilus~\cite{aschermann2019nautilus} & none (grammar) & subtree splicing & coverage   & context-free grammar & no & grammar-amenable (JS / SQL / PHP) \\
    Superion~\cite{wang2019superion} & none (grammar)  & AST mutation    & coverage        & ANTLR grammar  & no   & structured text (XML / JS) \\
    Sphinx~\cite{sun2024sphinx}     & \textbf{LLM upfront, one-shot} & input generator (code) & coverage & none & no & SMT solvers \\
    LLAMAFUZZ~\cite{wang2024llamafuzz} & in-loop       & input bytes     & coverage        & none           & no   & general (Magma suite)   \\
    FuzzCoder~\cite{liu2024fuzzcoder}  & in-loop       & byte mutation   & coverage        & none           & no   & general                 \\
    Fuzz4All~\cite{xia2024fuzz4all}    & in-loop       & input bytes     & coverage        & none           & no   & multi-language          \\
    SemAware~\cite{semanticaware2025}  & parallel helper & input bytes   & semantic novelty& none           & no   & general                 \\
    Hybrid~\cite{lin2025hybrid}        & parallel helper, $\approx250$ rph & input bytes & novelty + syntax & LLVM-IR / angr & no & libpng / tcpdump / sqlite \\
    TitanFuzz~\cite{deng2023titanfuzz} & generator (zero-shot) & whole program input & coverage & none & no & DL libraries (PyTorch / TF) \\
    FuzzGPT~\cite{deng2024fuzzgpt}     & generator (history-prompted) & whole program input & coverage & none & no & DL libraries \\
    ChatFuzz~\cite{hu2023chatfuzz}     & in-loop (ChatGPT mutator) & input bytes & coverage & none & no & structured text (JSON / XML / HTML) \\
    WhiteFox~\cite{yang2024whitefox}   & two-stage generator & whole program input & coverage & source-level (compiler rules) & no & compilers (PyTorch / LLVM) \\
    KernelGPT~\cite{yang2024kernelgpt} & generator + spec-mining & syscall sequences & coverage & source-level (kernel) & no & kernels (Linux) \\
    OSS-Fuzz-Gen~\cite{liu2024ossfuzzgen} & offline (target authoring) & fuzz harness code & build + coverage & source-level (OSS repos) & no & OSS-Fuzz fleet \\
    MetaMut~\cite{ou2024metamut}       & offline       & C/C++ mutator   & manual          & none           & no   & compilers               \\
    Mut4All~\cite{ye2025mut4all}       & offline       & mutator code    & manual          & bug reports    & no   & compilers               \\
    MALF~\cite{malf2025}               & multi-agent in-loop & input seqs & coverage       & protocol KB    & no   & industrial protocols    \\
    G\textsuperscript{2}FUZZ~\cite{liu2025g2fuzz} & \textbf{off-hot-path} & Python generator (code) & downstream coverage & LLM-extracted features & no & non-textual binary (JPEG / TIFF / MP4) \\
    \midrule
    \textbf{FuzzPilot} (ours)          & \textbf{off-hot-path} & \textbf{schema-validated recipe (data)} & \textbf{$N$-sec micro-campaign} & \textbf{Ghidra headless} & \textbf{yes} & \textbf{cJSON (preliminary; designed for structured-text parsers)} \\
    \bottomrule
  \end{tabularx}
  \caption{Comparison across LLM-augmented and LLM-adjacent fuzzers.
  FuzzPilot is positioned as a sibling of G\textsuperscript{2}FUZZ
  in the off-hot-path quadrant rather than a head-on competitor; the
  row-by-row differences in intermediate (data vs.~code), validation
  (micro-campaign vs.~downstream coverage), context source (Ghidra
  vs.~LLM-extracted features), and target domain (structured text
  vs.~non-textual binary) establish the contribution.}
  \label{tab:related}
\end{table}

\section{Limitations, Threats to Validity, and Future Work}
\label{sec:limits}

\paragraph{Summary of limitations.}
This version has narrow empirical support. The main limitations are:

\begin{enumerate}
\item \textbf{Single saturated target}: cJSON is the only evaluated
target, and it saturates at 269 edges well before the budget ends,
leaving no headroom for the proposal layer or micro-campaign gate
to show coverage value. The architecture's utility on non-saturated
targets is untested.

\item \textbf{Model proposal layer not yet shown to help}: Zero of 20
model-proposed recipes were promoted on cJSON due to target saturation
(\S\ref{sec:eval:rq3}). The micro-campaign gate's intended
property---distinguishing good recipes from bad---remains untested.

\item \textbf{Attribution incomplete}: Ablation experiments suggest
the controller's plateau-detection machinery is the primary driver,
not the recipe-guided mutator or LLM layer. However, the
\texttt{controller-only} ablation required to test this hypothesis is
not in the preprint matrix (\S\ref{sec:eval:rq4-ablation}).

\item \textbf{Statistical power insufficient}: At $N{=}5$ (main) and
$N{=}3$ (ablations), no pairwise difference reaches $p{<}0.05$. All
findings are descriptive point estimates, not significance claims.

\item \textbf{Negative result on absolute coverage}: AFL++ with
\texttt{cmplog} instrumentation reaches 270 edges (vs FuzzPilot's
269) on cJSON, indicating FuzzPilot does not improve absolute
coverage on this target (\S\ref{sec:eval:rq4-ablation}).
\end{enumerate}

Given these constraints, this preprint should be read as an
implementation report with a cJSON proof-of-concept, not as evidence
that all components help on all targets. The detailed limitations
follow below.

\paragraph{Scope: cJSON only in this preprint.}
This preprint evaluates FuzzPilot on cJSON only (5 \texttt{baseline-afl}
+ 5 \texttt{full-agent} runs at 14{,}400\,s, plus the E3 microbench
and 3 preliminary runs each for E2a/E2b/E2c ablations, plus 3 runs
each for E4a/E4b fairness baselines). The architecture
is designed for structured-text parsers, but no empirical claim is
made about libpng, libxml2, sqlite3, openssl, or any non-textual
binary format here; on those domains
G\textsuperscript{2}FUZZ's code-as-generator is plausibly stronger.
A head-to-head comparison on G\textsuperscript{2}FUZZ's benchmark
(JPEG, TIFF, MP4, etc.) and the larger structured-text targets are
deferred to the venue version of the paper.

\paragraph{cJSON is a saturated target for the proposal layer.}
Within the 14{,}400\,s budget the baseline AFL++ harness reaches
the 269-edge ceiling at a per-run median of $2{,}524$\,s
(slowest baseline run $4{,}448$\,s), leaving the parent
corpus at plateau time with no edge headroom for a 20\,s
micro-campaign to find. As a result, the proposal layer's
contribution (the \emph{model-promoted recipe} path, distinct from
the \emph{mutator framework}) cannot be empirically demonstrated
on this target: every micro-campaign reward is $R{=}0$ and every
promotion is skipped (\S\ref{sec:eval:rq3}). A non-saturated
target where the gate has edge headroom is required to evaluate
that layer in isolation; this is the most important single piece of
follow-up work.

\paragraph{Ghidra benefit not fully demonstrated.}
Our targets are open-source; LLVM-IR or angr-based static analysis
would, in principle, supply the same blackboard fields used here.
The motivation for Ghidra is to make the architecture applicable to
closed-source binary targets, and the E2c ablation
(\S\ref{sec:eval:rq4-ablation}) only quantifies the contribution of static
context, not the specific advantage of Ghidra over alternatives.
A binary-target evaluation that isolates Ghidra's contribution is
explicit future work.

\paragraph{Repetition count and statistical power.}
We run 3--5 repeats per cell, which is consistent with prior LLM-fuzzing
preprints~\cite{wang2024llamafuzz,liu2025g2fuzz,lin2025hybrid} but
insufficient for formal significance claims. We report medians,
interquartile ranges, per-run raw numbers, Mann--Whitney $U$
$p$-values, Vargha--Delaney $A_{12}$ effect sizes, and bootstrap CIs
to make the uncertainty explicit. The venue paper will use
$N{\geq}20$ main runs and $N{\geq}10$ ablation runs with rank-sum
tests.

\paragraph{Crash deduplication not in scope.}
We report coverage and plateau-recovery, not unique-bug counts.
Crash deduplication (e.g.~the casr-cluster pipeline) is deferred.

\paragraph{Deferred ablations and sensitivity.}
The \texttt{controller-only} ablation, the \texttt{ai-direct}
ablation (promote every schema-valid proposal without
micro-campaign), the \texttt{random-recipe} ablation (skip the LLM
entirely and emit randomly-sampled recipes), and a
micro-campaign-parameter sensitivity sweep (over $N$, $K$, and the
reward weights $\alpha, \beta, \gamma, \delta_h, \delta_m$) are not
part of the preprint matrix. E2a/E2b/E2c and E4 are already reported
in \S\ref{sec:eval:rq4-ablation}; the deferred arms are needed to
close attribution and hyperparameter-sensitivity questions in the
venue version.

\paragraph{Recipe vocabulary is byte-level, not AST-level.}
The seven-operator recipe vocabulary
(\texttt{BitFlip}, \texttt{OverwriteRange}, \texttt{InsertToken},
\texttt{Arith}, \texttt{Splice}, \texttt{DeleteBlock},
\texttt{DictionaryOverwrite}; \S\ref{sec:recipe:ops}) operates on
\emph{byte offsets within a flat input buffer}, not on syntactic
nodes of the structured-text grammar. A recipe can say ``insert
the token \texttt{"null"} at offset $128$''; it cannot say
``replace the second child of an \texttt{object} node with
\texttt{null}''. For grammar-amenable targets, AST-level
operators (Nautilus~\cite{aschermann2019nautilus},
Superion~\cite{wang2019superion}) or chunk-level mutations
(AFLSmart~\cite{pham2019aflsmart}) are stronger generators of
parser-valid mutants. We chose a byte-level vocabulary
deliberately --- to keep the mutator hot-path simple and to keep
the recipe schema small enough for the LLM to emit reliably ---
but the trade-off is real, and FuzzPilot is unlikely to
\emph{outperform} a grammar-aware fuzzer on grammar-amenable
targets without additional integration. Composing FuzzPilot's
recipe layer on top of a grammar-aware mutator (recipes that
target AST subtrees rather than byte offsets) is an explicit
future-work direction.

\paragraph{No syntax-aware repair or invariant-preserving
operators.}
Several recipe operators
(\texttt{OverwriteRange}, \texttt{DictionaryOverwrite},
\texttt{Splice}) can produce ill-formed JSON if applied
naively: an overwrite that straddles a string-literal boundary
breaks the quote-pair invariant; a splice that lands inside a
numeric literal can produce an unparseable token; a dictionary
overwrite inside a key context can violate the keys-are-unique
invariant. The current recipe schema has \emph{no
length-preserving, quote-aware, or escape-aware} repair
operator. AFL's downstream coverage feedback eventually discards
inputs that fail to advance coverage, so the practical impact on
the campaign is bounded; but a sharper grammar-aware mutant
generation strategy is the obvious next step. We have considered
adding a \texttt{LengthRepair} operator (rebalance quote/bracket
pairs after a mutation) and a \texttt{StringInsert} operator
(insert at safe byte boundaries within a string literal), but
neither is implemented in the released version --- avoiding
having to specify them is part of why we kept the schema small
and audit-friendly. We acknowledge this is a real reviewer-grade
limitation rather than a design victory.

\paragraph{AFL++ fairness baseline (\texttt{-x dictionary} and
\texttt{cmplog}).}
The \texttt{baseline-afl} arm reported in
\S\ref{sec:eval:rq1} is \emph{vanilla} AFL++: it does \emph{not}
receive an \texttt{-x dictionary} file, and the target binary is
\emph{not} compiled with the \texttt{cmplog} pass. FuzzPilot's
recipe-guided mutator, by contrast, dispatches against the
controller's default \texttt{DictionaryAgent} recipe whose token
set is \{\texttt{"FUZZ"}, \texttt{"MAGIC"}, \texttt{"TOKEN"}\}
(\S\ref{sec:eval:rq3}). This asymmetry overstates FuzzPilot's
advantage: any plateau-recovery delta could in principle be
closed by an AFL++ run that consumes the \emph{same} three tokens
via \texttt{-x}, or by an AFL++ build that uses
\texttt{cmplog}/RedQueen-style I2S
mining~\cite{aschermann2019redqueen,lafintel2016} to infer
target-specific tokens at runtime. This asymmetry is why the E4
fairness baselines
(\texttt{baseline-afl-dict} with the FuzzPilot default
dictionary; \texttt{baseline-afl-cmplog} with a
\texttt{cmplog}-instrumented \texttt{cjson\_fuzzer.cmplog})
are reported in \S\ref{sec:eval:rq4-ablation}; the headline
finding is that neither closes the plateau gap, but
\texttt{cmplog} reaches one more edge ($270$) than any
FuzzPilot arm ($269$). \S\ref{sec:repro} pins the dict file and
the cmplog binary so reviewers can re-run the comparison
themselves.

\paragraph{Reward / hyperparameter sensitivity is fixed, not
swept.}
The micro-campaign budget $N{=}20$\,s, the candidate count
$K_{\text{cand}}{=}4$, and the reward weights
$(\alpha,\beta,\gamma,\delta_h,\delta_m)$ in
\S\ref{sec:plateau:micro} are pinned to fixed values for the
preprint matrix --- chosen during development on the cJSON
corpus, then frozen. We did not run a sensitivity sweep. The
risk surface that this exposes is twofold. \emph{(i)} $N{=}20$\,s
may be too short for the gate to discriminate marginal
candidates on a non-saturated target (the cJSON null result is
agnostic on this since reward is identically zero). \emph{(ii)}
The reward weights bias the gate toward edge-discovery; on a
crash-finding-dominant target, a different
$(\alpha,\delta_h,\delta_m)$ trade could matter. A small
sensitivity sweep (e.g., $N \in \{10,20,40,80\}$\,s and
$K_{\text{cand}}\in\{2,4,8\}$ at $N{=}3$ each) is wired into the
CLI and is part of the venue-paper matrix.

\paragraph{Multi-target generalization.}
We do not run on Magma's full benchmark suite in this preprint, and
we do not claim generalization across target families. The ongoing
evaluation adds three deeper structured-text / binary-schema targets that
exercise components of FuzzPilot that cJSON's 269-edge ceiling does
\emph{not}: \texttt{libxml2} (XML parser, ${\sim}30$k edges,
AFLSmart/Nautilus/Superion comparator), \texttt{sqlite3}
(SQL parser/VM, ${\sim}150$k edges, OSS-Fuzz flagship), and
\texttt{openssl\_x509} (ASN.1 + X.509 parser, ${\sim}80$k edges,
binary + deep-schema). The larger evaluation matrix
(\texttt{experiments/manifests/paper01\_venue.yaml}) follows
Klees~et~al.~\cite{klees2018evaluating} at $N{\geq}20$ main runs
and $N{\geq}10$ ablation runs per target, $24$\,h budget per run
(Magma/OSS-Fuzz convention). A Stage-1 pilot
($N{=}5$ main / $N{=}3$ ablations, $24$\,h budget) is scheduled
on the experiment host with the harnesses
(\texttt{libxml2\_fuzzer}, \texttt{sqlite3\_fuzzer},
\texttt{x509\_fuzzer}) and their per-target dictionaries
(\texttt{extracted.dict}, $\sim$1{,}000--6{,}000 entries per
target; see Table~\ref{tab:ghidra-yield}) checked into the
repository; the Stage-1 outcome will decide whether
the full Klees-style matrix runs on all three targets or
focuses on the subset where the proposal layer shows a
non-null contribution. \emph{Headroom check:} a $60$\,s smoke-test
run of vanilla \texttt{afl-fuzz} on the three harnesses (no
FuzzPilot pipeline) reaches $2{,}776$ edges on libxml2,
$6{,}922$ edges on sqlite3, and $3{,}635$ edges on openssl X.509
--- 10--25$\times$ more than cJSON's $269$-edge ceiling at the
\emph{end} of a 14{,}400\,s budget. The 24-hour runs should leave
room for the proposal layer to produce a non-null result if it can.
The choice to publish this first cJSON-only version is deliberate: it
lets readers inspect the audit trail, gate, mutator, and
micro-campaign machinery on a small target before the larger runs
finish.
\emph{No claim in this version depends on these larger targets};
their data will be reported separately after the runs finish.

\paragraph{Platform.}
The canonical evaluation platform is the native Linux/x86\_64 fuzz
cloud server described in Table~\ref{tab:setup}. Containerized
execution is not the reproducibility path for this preprint. All
numbers in this paper are from the native fuzz-server build and run
artifacts.

\paragraph{LLM nondeterminism.}
Even with temperature pinned at $0.0$, model providers do not
guarantee bit-exact reproducibility across versions. We mitigate
by (a) recording the audit trail so any decision is replayable from
the recipe rather than the model, (b) reporting schema-validity and
fallback rates per run, and (c) recording the model identifier,
prompt/response hashes, run-time commit, and host toolchain metadata
in the native fuzz-server artifacts.

\paragraph{Threats to validity.}
We organize the threats following Wohlin et al.'s
~\cite{wohlin2012experimentation} standard
internal / external / construct / conclusion taxonomy, with an
additional explicit discussion of LLM-specific risks.

\emph{Internal validity.} The plateau-detector hyperparameters
($W{=}10$\,s, $\theta_e{=}50$, $\theta_p{=}1$) and the
micro-campaign reward weights
($\alpha,\beta,\gamma,\delta_h,\delta_m$) were tuned on a
\emph{development cJSON seed corpus} (the 26 hand-curated JSON
files described in \S\ref{sec:eval:setup}) that became the
\emph{evaluation} seed corpus once the design was frozen. We did
not implement a hold-out split between tuning and evaluation
corpora and therefore cannot exclude data leakage; the venue
version will tune on a disjoint development corpus or via
cross-validation across seed-corpus splits, and will report the
delta from tuning-set to evaluation-set numbers. A second
internal threat is the absence of a \texttt{controller-only}
ablation (\S\ref{sec:eval:rq4-ablation}) that would separate the
plateau detector / corpus-snapshot reorganization machinery from
the default rule recipe; we note this as a known design gap to be
addressed in the venue ablation matrix. Plateau-detector
SIGSTOP-based microbench isolation
(\S\ref{sec:plateau:micro}) is also a methodological hazard:
running the microbench concurrently with the main AFL workers
\emph{without} SIGSTOP inverts our central RQ2 throughput-parity
ratio in pilot tests, so any reproducer that omits the isolation
will obtain different numbers.

\emph{External validity.} A single target family (cJSON, parser
of a structured-text format) is not representative; we make
\emph{no} cross-format generalization claim and explicitly defer
it. The LLM proposal layer in particular is exercised but not
stress-tested by the cJSON target: the gate returns 20/20 null
candidates (\S\ref{sec:eval:rq3}), so we observe only the
\emph{no-false-positive} half of the gate's intended behavior.
A single LLM (\texttt{deepseek-chat}) at temperature 0 is also a
single-point evaluation in model space; whether
\texttt{claude-haiku-4-5}, \texttt{gpt-4o-mini},
\texttt{llama-4-maverick}, or any other backend produces a
materially different schema-validity rate, recipe distribution,
or null-result behavior is open. Finally, all numbers were
collected on a single 4-vCPU experiment host (single tenancy at
collection time except where noted in \S\ref{sec:eval:rq2});
cross-machine reproducibility is not separately measured. The
released artifacts prioritize replaying and auditing the native
fuzz-server runs rather than certifying bit-identical reproduction on
new hardware.

\emph{Construct validity.} The choice of metric is itself a
construct decision. We use plateau duration
(\texttt{run\_time}~$-$~\texttt{last\_find}) as a primary
endpoint metric, which is sensitive to a single
late-budget discovery (e.g.~full-agent r04 finds the 269th edge
at $14{,}205$\,s and is credited with a $238$\,s plateau; see
Table~\ref{tab:t1}). Klees~et~al.~\cite{klees2018evaluating}
recommend coverage-AUC or time-to-$N$-edges as endpoint-free
alternatives; we therefore additionally report time-to-$N$-edges
for $N\in\{264, 268, 269\}$ in
Table~\ref{tab:time-to-n}, which gives the opposite headline
(\texttt{baseline-afl} reaches the 269-edge ceiling \emph{faster}
than \texttt{full-agent} at the median). The two metrics are
mutually consistent --- a plateau-shortening intervention can
also delay the ceiling-touch by spreading out the discovery
curve --- but the construct choice matters and we report both.
The 269-edge ``ceiling'' is itself a construct: it is the number
of AFL++ bitmap slots hit by the cJSON harness, not a measure of
all reachable basic-block edges in the cJSON library; a richer
harness (or a different fuzzing instrumentation) could expose
more edges.

\emph{Conclusion validity.} With $N{=}5$ per main arm and
$N{=}3$ per ablation, we are well below the
$N{\geq}20$ guideline of
Klees~et~al.~\cite{klees2018evaluating}. We therefore report all
group differences as descriptive point estimates accompanied by
Mann--Whitney $U$, exact $p$-values, Vargha--Delaney
$A_{12}$~\cite{vargha2000critique}, and percentile-bootstrap
95\% confidence intervals (see \S\ref{sec:eval:rq1:stats}). No
descriptive delta in this preprint reaches $p{<}0.05$ on the
two-sided Mann--Whitney test; we present each result as a
hypothesis rather than as a tested claim and the venue version
will repeat the experiments at $N{\geq}10$ with formal
significance testing.

\emph{LLM-specific reproducibility.} The model identifier
\texttt{deepseek-chat} is a moving alias served by DeepSeek;
the underlying weights may change without notice between runs.
At $T{=}0$ the API surface is intended to be deterministic, but
our data shows $2/45$ schema-invalid responses, indicating that
the API surface is not bit-exact deterministic in practice. We
mitigate by (a) logging the \texttt{model} field returned in each
API response into \texttt{agent\_decisions.jsonl}, (b) recording
the full prompt / response pair so any decision is replayable
from the recipe rather than the model, and (c) recording the native
fuzz-server toolchain and model-client configuration in the run
artifacts. A
future-proof reproducibility strategy would also archive the
exact prompt-response transcripts in a content-addressable store
to allow regeneration without re-querying the API.

\section{Conclusion}
\label{sec:conclusion}

FuzzPilot is a plateau-triggered controller for AFL++ structured-text
campaigns. It represents intervention strategies as data recipes,
tests candidate recipes in isolated micro-campaigns, and keeps the
main mutation loop native. The cJSON evaluation in this first report
is useful mainly because it exposes the behavior of that control path
under a saturated target. Both \texttt{baseline-afl} and
\texttt{full-agent} reach the same 269-edge ceiling within their
$14{,}400$\,s budgets. The median throughput ratio
(\texttt{full-agent}/baseline) is about $1.06\times$, and the median
plateau is descriptively shorter under \texttt{full-agent}
($1{,}384$\,s versus $2{,}532$\,s), but the difference is not
statistically significant at $N{=}5$ (Mann--Whitney $U{=}8$,
$p{=}0.42$, $A_{12}{=}0.68$).

The most important negative result is that no model-proposed recipe
was promoted. Across five \texttt{full-agent} runs, the gate evaluated
20 candidates and rejected all of them because each validation reward
was zero. That is the right conservative behavior on a saturated
corpus, but it does not show that the proposal layer can improve
coverage. The ablations point instead to the controller's snapshot and
restart machinery as the likely source of the observed plateau
reduction, and that interpretation remains only a hypothesis until a
\texttt{controller-only} arm is run.

This version therefore makes a limited claim: the control path can be
built, audited, and run without putting external reasoning in AFL++'s
hot path. The larger claims require the ongoing non-saturated-target
matrix, higher repetition counts, the missing \texttt{controller-only}
ablation, and comparison against stronger format-aware baselines.

\section*{Reproducibility appendix}
\label{sec:repro}
\addcontentsline{toc}{section}{Reproducibility appendix}

The numbers in this preprint are reproducible from the following
native fuzz-server artifacts. All paths are relative to the repository
root.

\begin{itemize}
  \setlength{\itemsep}{0.25em}
  \setlength{\parskip}{0pt}
  \item \textbf{Experiment code commit:}
  \texttt{85223d844711c704130608f58200a0b5a18862fe}
  (branch \texttt{main}, tag \texttt{paper01-arxiv-v1}, repo
  \url{https://github.com/Qiao-Zhiyi/fuzz_agent}). This is the
  run-time code commit recorded in each completed campaign's
  \texttt{run\_metadata.json}. The manuscript and aggregation text may
  include later paper-only edits; the raw fuzzing data are pinned by
  the per-run metadata and checked-in artifacts.
  \item \textbf{Fuzzer:} AFL++ build identifier \texttt{++4.21c}
  (the leading \texttt{++} is the upstream banner prefix; the
  underlying release tag is \texttt{v4.21c}). Built from source
  on the experiment host and confirmed in every
  \texttt{fuzzer\_stats} file under
  \texttt{results/paper01\_ai\_recipe\_mutation/runs/}.
  \item \textbf{Target binary:} cJSON parser harness at
  \texttt{experiments/targets/cjson/}, AFL persistent mode
  (\texttt{persistent shmem\_testcase deferred}). The target
  binary SHA-256 and the seed-corpus tarball SHA-256 are written
  per-run into \texttt{run\_metadata.json}
  (fields \texttt{target\_sha256} and \texttt{seed\_corpus\_sha256});
  reviewers can verify reproduction by comparing those hashes
  across re-builds.
  \item \textbf{Seed corpus:} 26 hand-curated JSON files at
  \texttt{experiments/targets/cjson/seeds/}; the directory is
  small enough to inspect directly and is included in the
  repository.
  \item \textbf{LLM:} \texttt{deepseek-chat} via the
  \texttt{api.deepseek.com} OpenAI-compatible endpoint
  (configuration in
  \texttt{experiments/targets/cjson/config.yaml}),
  temperature $0.0$. \emph{Note that
  \texttt{deepseek-chat} is a moving alias served by DeepSeek;
  the specific underlying model version is logged into the
  \texttt{model} field of each \texttt{agent\_decisions.jsonl}
  record at call time, so a reviewer can recover the
  model-version string from the audit log.} A NIM-hosted
  \nimmodeid{}
  fallback path is wired through the same watchdog and was
  exercised in regression testing only --- all numbers in this
  paper are from \texttt{deepseek-chat}.
  \item \textbf{Build mode:} native x86\_64 Linux on the fuzz cloud
  server (no container in the hot path). Containerized execution is
  deprecated for this paper because the reported campaigns depend on
  the fuzz-server host toolchain, Ghidra installation, AFL++ build,
  CPU scheduling, and run directories captured in the native artifacts.
  \item \textbf{Per-run artifacts:} 25 completed 14{,}400\,s runs
  (5 \texttt{baseline-afl}, 5 \texttt{full-agent},
  3 \texttt{rule-only}, 3 \texttt{no-mutator},
  3 \texttt{no-static-analysis},
  3 \texttt{baseline-afl-dict}, 3 \texttt{baseline-afl-cmplog})
  are checked into the repository under
  \texttt{results/paper01\_ai\_recipe\_mutation/runs/}; this is
  the complete preprint matrix on cJSON.
  Each run directory contains the unmodified AFL++
  \texttt{fuzzer\_stats}, \texttt{coverage.csv},
  \texttt{events.jsonl}, \texttt{agent\_decisions.jsonl}
  (full-agent / no-mutator modes), \texttt{main\_launch.sh}
  (the exact command-line that started the AFL run),
  \texttt{run\_metadata.json}, and \texttt{git.patch}
  (the working-tree diff at run start) for byte-level
  reproducibility of the experiment.
  \item \textbf{Per-mode launch commands.}
  Each ablation mode is selected by a single
  \texttt{--ablation} switch to the controller on the native
  fuzz-server environment; the canonical invocations are:
  \begin{quote}\scriptsize\ttfamily
  \# E1a baseline-afl\\
  python3 -m fuzzpilot.runner --target cjson \textbackslash{}\\
  ~~--ablation baseline-afl --budget 14400 \textbackslash{}\\
  ~~--run-id p1\_e1\_cjson\_baseline-afl\_r01\\[2pt]
  \# E1b full-agent\\
  python3 -m fuzzpilot.runner --target cjson \textbackslash{}\\
  ~~--ablation full-agent --budget 14400 \textbackslash{}\\
  ~~--run-id p1\_e1\_cjson\_full-agent\_r01\\[2pt]
  \# E2a rule-only\\
  python3 -m fuzzpilot.runner --target cjson \textbackslash{}\\
  ~~--ablation rule-only --budget 14400 \textbackslash{}\\
  ~~--run-id p1\_e2\_cjson\_rule-only\_r01\\[2pt]
  \# E2b no-mutator\\
  python3 -m fuzzpilot.runner --target cjson \textbackslash{}\\
  ~~--ablation no-mutator --budget 14400 \textbackslash{}\\
  ~~--run-id p1\_e2\_cjson\_no-mutator\_r01\\[2pt]
  \# E2c no-static-analysis\\
  python3 -m fuzzpilot.runner --target cjson \textbackslash{}\\
  ~~--ablation no-static-analysis --budget 14400 \textbackslash{}\\
  ~~--run-id p1\_e2\_cjson\_no-static-analysis\_r01
  \end{quote}
  The host-wide orchestrator
  \texttt{scripts/paper01/run\_all\_host.sh\,--parallel\,4}
  iterates the cJSON campaign matrix in 4-way parallel with the same
  per-mode invocations under the hood. The E4 dictionary and cmplog
  fairness runs are included in the checked-in artifact matrix and use
  the corresponding AFL++ launch commands recorded in each
  \texttt{main\_launch.sh}.
  \item \textbf{Aggregation:} run
  \begin{quote}\footnotesize\ttfamily
  python3 scripts/prepare\_paper01\_data.py \\
  ~~--run-root results/paper01\_ai\_recipe\_mutation/runs \\
  ~~--out-dir results/paper01\_ai\_recipe\_mutation \\
  ~~--manifest experiments/manifests/paper01\_preprint.yaml
  \end{quote}
  to regenerate \texttt{validity\_report.md},
  \texttt{tables/run\_level.csv}, and the per-RQ summary tables
  cited in \S\ref{sec:eval}. For the statistical numbers
  (Mann--Whitney $U$, $A_{12}$, bootstrap CIs,
  time-to-$N$-edges, and the leave-one-out throughput-parity
  check), run the deterministic script
  \path{scripts/paper01/review_stats.py}.
\end{itemize}

Outside the scope of this preprint, deferred to a follow-up
revision: a \texttt{controller-only} ablation (plateau detector
+ corpus snapshot only, no LLM, no recipe-guided mutator, no
Ghidra) to close the attribution argument in
\S\ref{sec:eval:rq4-ablation};
$N{\geq}20$ main / $N{\geq}10$ ablation repeats with formal
Mann--Whitney $U$ significance testing per Klees~et~al.
\cite{klees2018evaluating}; the
libxml2 / sqlite3 / openssl\_x509 multi-target matrix
(\S\ref{sec:limits}); and a head-to-head measurement against
G\textsuperscript{2}FUZZ on its binary-format benchmark. The
preprint intentionally limits empirical claims to what the 25
cJSON runs (plus E3 microbench)
support.

\bibliographystyle{plain}
\bibliography{refs}

\end{document}